\documentclass[twocolumn]{aa}
\usepackage{graphicx}
\usepackage{latexsym,natbib,graphicx}
\def\cf{cf.~}
\def\eg{e.g.~}
\def\fig{Fig.\,}
\def\sec{Sect.\,}
\def\eq{\!=\!}
\def\ltsim{~\rlap{\lower -0.5ex\hbox{$<$}}{\lower 0.5ex\hbox{$\sim\,$}}}
\def\gtsim{~\rlap{\lower -0.5ex\hbox{$>$}}{\lower 0.5ex\hbox{$\sim\,$}}}
\def\deg{\hbox{$^\circ$}\,}

\def\magsqarcsec{mag$/$\raisebox{-0.4ex}{\hbox{$\Box^{\prime\prime}$}\,}}
\def\s0{S0}
\def\kms{km\,s$^{-1}$}
\def\zk{\hbox{$z_0^{\rm k}$}\,}
\def\hk{\hbox{$h^{\rm k}$}\,}
\def\zn{\hbox{$z_0^{\rm n}$}\,}
\def\hn{\hbox{$h^{\rm n}$}\,}
\def\zkdzn{\hbox{$z_0^{\rm k}/z_0^{\rm n}$}\,}
\def\hkdhn{\hbox{$h^{\rm k}/h^{\rm n}$}\,}
\def\muk{\hbox{$\mu_0^{\rm k}$}\,}
\def\mun{\hbox{$\mu_0^{\rm n}$}\,}
\def\rco{\hbox{$R_{\rm co}$}\,}
\def\hin{\hbox{$h_{\rm in}$}\,}
\def\hout{\hbox{$h_{\rm out}$}\,}
\def\hindout{\hbox{$h_{\rm in}/h_{\rm out}$}\,}
\begin{document}
   \title{ Thick Disks of Lenticular Galaxies \thanks{Based on observations 
       obtained at the European Southern Observatory, Chile }
         }
   \subtitle{3D-photometric thin/thick disk decomposition of eight edge-on 
     \s0 galaxies}

   \author{M. Pohlen
          \inst{1,2}
          \and
          M. Balcells \inst{1}
          \and 
          R.~L\"utticke \inst{3,2}
          \and 
          R.-J. Dettmar \inst{2}
          }

   \offprints{M. Pohlen}

\institute{
Instituto de Astrof\'{\i}sica de Canarias, 
E-38200 La Laguna, Tenerife, Spain\\
\email{pohlen,balcells@ll.iac.es}
\and
Astronomical Institute, Ruhr-Universit\"at Bochum, 
D-44780 Bochum, Germany\\ 
\email{dettmar@astro.rub.de}
\and
Department of Computer Science, FernUniversit\"at Hagen,
D-58084 Hagen, Germany \\
\email{Rainer.Luetticke@FernUni-Hagen.de}
         }
   \date{Received 23 December, 2003; accepted 04 March, 2004}
   \abstract{ 
Thick disks are faint and extended stellar components found around several
disk galaxies including our Milky Way. The Milky Way thick disk, the only one 
studied in detail, contains mostly old disk stars ($\approx\!10$\,Gyr), 
so that thick disks are likely to trace the early stages of disk evolution.
Previous detections of thick disk stellar light in external galaxies have 
been originally made for early-type, edge-on galaxies but detailed 
2D thick/thin disk decompositions 
have been reported for only a scant handful of mostly late-type disk galaxies. 
We present in this paper for the first time explicit 3D thick/thin disk 
decompositions characterising the presence and 
properties (\eg scalelength and scaleheight) for a sample of eight 
lenticular galaxies by fitting 3D disk models to the data. 
For six out of the eight galaxies we were able to derive a consistent 
thin/thick disk model. The mean scaleheight of the thick disk is 3.6 times
larger than that of the thin disk. The scalelength of the thick disk is 
about twice, and its central luminosity density between 3-10\% of, the 
thin disk value. 
Both thin and thick disk are truncated at similar radii. This implies 
that thick disks extend over fewer scalelengths than thin disks, and 
turning a thin disk into a thick one requires therefore vertical 
but little radial heating. 
All these structural parameters are similar to thick disk parameters
for later Hubble-type galaxies previously studied. 
We discuss our data in respect to present models for the origin of 
thick disks, either as pre- or post-thin-disk structures, providing 
new observational constraints.
   \keywords{
Galaxies: photometry  -- 
Galaxies: bulges --
Galaxies: structure -- 
Galaxies: fundamental parameters  -- 
Galaxies: evolution  -- 
Galaxies: formation  -- 
Galaxies: individual: 
ESO311-012, NGC1596, NGC2310, NGC3564, NGC3957, NGC4179, NGC4521, NGC5047
            }                 
   }                          
   \maketitle                 
%
\section{Introduction}
\label{introduction}
\begin{table*}
\begin{center}
{\normalsize
\begin{tabular}{ l  c c l  r@{.}l  c  c  c r@{.}l c }
\hline
\rule[+0.4cm]{0mm}{0.0cm}
Galaxy
&RA 
&DEC
&RC3
&\multicolumn{2}{c}{T}
&Diam.
&$v_{\sun}$
&$v_{\rm vir}$
&\multicolumn{2}{c}{D} 
&$M^{\scriptscriptstyle 0}_{\rm B}$ \\[+0.1cm]
&\multicolumn{2}{c}{(2000.0)}
&type 
&\multicolumn{2}{c}{}
&[\ \arcmin\ ]
&[\ \kms]
&[\ \kms]
&\multicolumn{2}{c}{[Mpc]} 
&[mag]\\
\rule[-3mm]{0mm}{5mm}{\scriptsize{\raisebox{-0.7ex}{\it (1)}}}
&{\scriptsize{\raisebox{-0.7ex}{\it (2)}}}
&{\scriptsize{\raisebox{-0.7ex}{\it (3)}}}
&\hspace*{0.1cm}{\scriptsize{\raisebox{-0.7ex}{\it (4)}}}
&\multicolumn{2}{c}{{\scriptsize{\raisebox{-0.7ex}{\it (5)}}}}
&{\scriptsize{\raisebox{-0.7ex}{\it (6)}}}
&{\scriptsize{\raisebox{-0.7ex}{\it (7)}}}
&{\scriptsize{\raisebox{-0.7ex}{\it (8)}}}
&\multicolumn{2}{c}{{\scriptsize{\raisebox{-0.7ex}{\it (9)}}}} 
&{\scriptsize{\raisebox{-0.7ex}{\it (10)}}} \\
\hline\hline \\[-0.2cm]
NGC\,1596   &04 27 38.1&$-$55 01 40 &.LA..*/ &$-$2&0 &3.9 &1510 &1229   &17&1&$-$19.3\\
NGC\,2310   &06 53 54.0&$-$40 51 45 &.L..../ &$-$2&0 &4.1 &1187 & 943   &13&1&$-$18.6\\
ESO\,311-012&07 47 34.1&$-$41 27 08 &.S..0?/ &   0&0 &3.7 &1131 & 893   &12&4&$-$20.0\\
NGC\,3564   &11 10 36.4&$-$37 32 51 &.L...*/ &$-$2&0 &1.7 &2812 &2639   &36&7&$-$20.2\\
NGC\,3957   &11 54 01.5&$-$19 34 08 &.LA.+*/ &$-$1&0 &3.1 &1687 &1607   &22&3&$-$19.0\\
NGC\,4179   &12 12 52.6&$+$01 17 47 &.L..../ &$-$2&0 &4.1 &1248 &1277   &17&7&$-$19.6\\
NGC\,4521   &12 32 47.7&$+$63 56 21 &.S..0.. &   0&0 &2.6 &2500 &2767   &38&4&$-$20.2\\
NGC\,5047   &13 15 48.5&$-$16 31 08 &.L..../ &$-$2&0 &2.7 &6330 &6292   &87&4&$-$21.7\\
\hline
\end{tabular}
}
\caption[]{Global parameters of the observed lenticular galaxies: 
{\scriptsize{\it (1)}} Principal name, {\scriptsize{\it (2)}} right 
ascension, {\scriptsize{\it (3)}} declination,  {\scriptsize{\it(4)}} 
RC3 coded Hubble-type, and the {\scriptsize{\it(5)}} Hubble parameter T 
are taken from \cite{rc3}. The {\scriptsize{\it(6)}} diameter in arcminutes, the 
{\scriptsize{\it(7)}} heliocentric radial velocities, and the B-Band absolute magnitude
{\scriptsize{\it(10)}} are taken from  LEDA. 
According to the heliocentric radial velocities corrected for the Local Group 
infall into the Virgo cluster {\scriptsize{\it(8)}} from LEDA, we estimated the 
{\scriptsize{\it (9)}} distances following the Hubble relation with the 
Hubble constant from the HST key project 
of $H_{0}\!=\!72$ km s$^{-1}$Mpc$^{-1}$ \cite[]{hst_h0}.
\label{sample} }
\end{center}
\end{table*}
The knowledge of the detailed distribution of stars in galaxies  
is of fundamental importance to address the formation and evolution 
of those systems. 
To a first approximation, a disk galaxy can be described by a 
set of distinct stellar entities: a disk population, a bulge component, 
and a stellar halo. 
Deep surface photometry of external early-type galaxies 
\cite[]{bursteinI,bursteinII,bursteinIII,tsikoudi1979,tsikoudi1980} and 
later elaborate measurements in our own Galaxy \cite[]{gilmore1983} 
revealed the need for an additional component of stars. This was 
called `thick disk' \cite[]{bursteinIII}, since it exhibiting a disk-like 
distribution with larger scaleheight compared to the inner, 
dominating `thin disk'. 
There are three distinct families of hypotheses for 
its creation. The first group considers the thick disk as a separate entity 
produced in an early phase of enhanced star formation during the initial 
proto-galactic collapse 
\cite[i.e.~the ELS scenario,][]{els,gilmore1984,burkert1992}. 
Another family of models regards the thick disk as an extension (by dynamical 
heating) of the thin disk. They assume that after the initial collapse 
all gas settles down into the galactic plane and starts forming stars. 
On this thin stellar disk a variety of constant or violent heating 
mechanisms could act: spiral density waves \cite[]{barbanis1967,carlberg1985}, 
encounters with giant molecular clouds \cite[e.g.][]{spitzer1953,lacey1984}, 
scattering by massive black holes \cite[]{lacey1985}, energy input by 
accretion of satellite galaxies \cite[]{carney1989,quinn1993,velazquez1999,
aguerri2001}, or bar bending instabilities \cite[]{raha1991}.
For example, \cite{gnedin2003} recently used N-body simulations to 
show that tidal heating in a cluster is sufficient to thicken 
stellar disks by a factor of 2-3. This kinematic heating
and vertical expansion will lead to a significant morphological 
transformation of a normal spiral galaxy into a lenticular.   
The third model suggests that thick disks are mostly made of 
debris material from accreted satellites. Recent cosmological 
N-body+SPH galaxy formation models of \cite{abadi2003} locating the 
thick disk formation before $z\!\approx\!1$ find that more than 
half of the thick disk stars are actually tidal debris from 
disrupted satellites. Therefore the thick disk is not a former 
thin disk thickened by a minor merger.
To decide which of these hypotheses could explain the thick disk 
phenomenon best we need first a more general and complete statistic 
of thick disk properties. Naturally, these are rather global ones 
for external galaxies whereas our particular position in the Milky Way 
makes it possible to determine much finer details.
Since the work of Tsikoudi and Burstein \cite[]{bursteinI,bursteinII,
bursteinIII,tsikoudi1979,tsikoudi1980} it appears well known that 
thick disks are quite common in \s0 galaxies. 
However, none of the more recent detections, except for two galaxies 
in \cite{degrijs1996} and a short remark in \cite{degrijs1997}, quantifying 
detailed parameters such as the ratio of thick to thin disk scaleheight or 
scalelength, is actually made in \s0 galaxies.
All these galaxies are of later Hubble type. In addition, we have not 
found a detailed 2D thin/thick disk decomposition
for any \s0 galaxy in the literature. 
Subsequent numerical decompositions dealing with \s0 galaxies after 
the pioneering work in the early 80's treated the thick disk either 
as an outer flattened but exponential 
halo \cite[for NGC\,4452 and NGC\,4762:][]{hamabe1989}, or as 
a spheroidal bulge component 
\cite[for NGC\,1381:][]{decarvalho1987} 
\cite[for NGC\,3115:][]{capaccioli1987,silva1989}.
The detections of possible halo or thick disk stellar light in 
disk galaxies of later Hubble type have been made in a scant 
handful of mostly nearby edge-on galaxies 
\cite[\eg for ESO\,342-017, IC\,5249, NGC\,891, NGC\,4565, NGC\,5907, 
and NGC\,6504:][]{vdk81,vdk81b,shaw1989,morrison1994,vandokkum1994,sackett1994,
morrison1997,naeslund1997,lequeux1998,zheng1999,neeser2002}.
Quite recently, \cite{dalcanton2002} suggest the detection 
of extended, ubiquitous thick disks in a large sample of 
late-type, edge-on galaxies by means of multi-colour imaging. 
However, their thick disks are solely detected by vertical 
colour gradients for which dust extinction complicates 
the interpretation. 
In addition, their vertical colour profiles, especially $(R-K)$, 
typically extend out to only very few vertical disk scaleheights. 
At those z-heights where they attribute the red colour to an additional 
component the thin disk may still be dominant over a potential thick 
disk and even determine the measured colour.  
In this paper, we analyse a set of eight edge-on 
\s0 galaxies using the classical approach for the identification of
thick disks in external galaxies: the need for an additional disk component 
when attempting to fit single disk models to the light distribution 
in a deep image.
Thereby we characterise the presence and properties (scalelengths, -heights, 
and central surface brightnesses) of thick disks by directly measuring 
their structure. 
%
\section{Sample Selection \& Observations}
\label{data}
The data were taken as part of the PhD study on the radial structure on 
galactic stellar disks by \cite{pohlen2001}. The galaxies were selected 
according to the allocated observing time, observatory, and CCD-chip 
size meeting the following morphological selection criteria. 
Using images from the Digitized Sky Survey (DSS) we verified 
that they are edge-on ($i\gtsim 86$\deg), undisturbed, 
and similar to some disk-prototypical cases like NGC\,4565 
or IC\,2531 to make it possible to consistently fit the applied 
simple disk model 
(\cf\sec\ref{model}).
Galaxies with the following characteristics were rejected: spiral arms 
(indicating towards a lower inclination), a significantly asymmetric or 
disturbed disk (indicating towards strong interaction), and two-sided or 
significantly one-sided warped disks. In addition, galaxies which 
seemed to be dominated by the light of the bulge component, or appeared
to be too small ($D_{25}\!\gtsim\!2$\arcsec), or showed only a faint, 
patchy disk were also excluded. 
For the observed lenticulars we therefore selected galaxies 
with a symmetric, smooth disk and a distinct, non-dominant 
bulge component. It is not unexpected that half of these galaxies 
show a box- or peanut-shaped (b/p) bulge component, since \cite{luett2000a} 
find that $>\!40$\% of all bulges are b/p shaped. 
Global properties of the finally observed eight \s0 galaxies are given in
Table~\ref{sample}. 
We do not argue that this sample is fully representative of the general 
population of \s0 galaxies but it ensures the best prospects for obtaining
consistent models with our 3D modelling technique for all galaxies.   
The images (in Johnson R or V filter) were obtained in four observing runs 
in 1998/1999, three at the Danish 1.54m telescope of the European Southern 
Observatory (ESO, Chile) and one at the 1.23m telescope 
on Calar Alto (CAHA, Spain). 
During all three runs at the ESO the 1.54m Danish telescope was equipped 
with DFOSC and the C1W7/CCD which is a 2k\,x\,2k LORAL chip providing a 
field size of $\approx\!13$\arcmin\ and a scale of $\approx\!0.39$\arcsec\ 
pixel$^{-1}$. 
The run at the Calar Alto 1.23\,m telescope was done in service mode with 
the Site\#18b chip, a 2048x2048 SITE CCD with 24\,$\mu$m 
pixel size, providing an unvignetted field of $\approx 10$\arcmin\  and a 
scale of $\approx 0.5$\arcsec\ pixel$^{-1}$.
The standard CCD reduction techniques 
---overscan correction; subtraction of remaining large scale gradient 
in combined, oversan-subtracted, masterbias image; and careful 
flatfielding-- were applied using the IRAF data reduction package.
Neither the DFOSC nor the Calar Alto CCD R-band images were affected 
by fringing. 
The individual, dithered, reduced short exposures (150\,s-600\,s) 
were combined to the final deep image using IRAF's {\sl imcombine} task. 
These images are rotated to the major axis using the smallest
angle of rotation according to their true position on the sky.
Table \ref{observ} summarises the detailed observational parameters.
\begin{table*}
\begin{center}
\begin{tabular}{l c c c c c c }
\hline
\rule[+0.3cm]{0mm}{0.0cm}
Galaxy
&Filter 
&Date
&Site
&Exp.time
&Seeing
&Coadds \\
 &&[mmyy]&&[min]&[\arcsec]&\#\,x\,[s] \\[+0.1cm]
\rule[-3mm]{0mm}{5mm}{
\scriptsize{\raisebox{-0.7ex}{\it (1)}}}
&{\scriptsize{\raisebox{-0.7ex}{\it (2)}}}
&{\scriptsize{\raisebox{-0.7ex}{\it (3)}}}
&{\scriptsize{\raisebox{-0.7ex}{\it (4)}}}
&{\scriptsize{\raisebox{-0.7ex}{\it (5)}}}
&{\scriptsize{\raisebox{-0.7ex}{\it (6)}}}
&{\scriptsize{\raisebox{-0.7ex}{\it (7)}}} \\
\hline\hline \\[-0.2cm]
NGC\,1596   & R & 0198 & ESO & 45.0& 1.3  &8x300, 2x150  \\
NGC\,2310   & R & 0198 & ESO & 51.7& 1.3  &600, 500, 200, 9x200 \\
ESO\,311-012& V & 0599 & ESO & 60.0& 1.6  &12x300 \\
NGC\,3564   & V & 0399 & ESO & 66.3& 1.3  &600, 500, 3x360, 6x300 \\ 
NGC\,3957   & R & 0198 & ESO & 30.0& 1.3  &3x600 \\
NGC\,4179   & V & 0599 & ESO & 65.0& 1.6  &1x600, 1x480, 1x360, 1x300, 9x240 \\
NGC\,4521   & R & 0699 & CAHA& 60.0& 1.5  &6x600 \\
NGC\,5047   & V & 0599 & ESO & 60.0& 1.5  &6x600 \\
\hline \\
\end{tabular}
\caption{Observing log for the individual combined images with
{\scriptsize{\it (1)}} the galaxy, {\scriptsize{\it (2)}} the 
filter, {\scriptsize{\it (3)}} the observing date, {\scriptsize{\it (4)}} 
the site, {\scriptsize{\it (5)}} the total coadded on-source 
exposure time $t_{\rm int}$, {\scriptsize{\it (5)}} mean seeing conditions 
during the observations, {\scriptsize{\it (7)}} the number of individual 
images with their $t_{\rm int}$.
}
\end{center}
\label{observ}
\end{table*}
During the two ESO observing runs in 1999 several \cite{landolt} fields, 
partly enriched with additional stars provided by B.~Skiff 
(priv.~comm.), were observed. 
The standard fields were taken at least three times a night at different 
airmasses to determine the atmospheric extinction. 
During the other two observing runs no standard stars were taken.
The ESO run in 1998 is calibrated by literature values and for galaxies 
without catalogued values interpolated according to the measured 
sky background.
Only a rough zero point could be estimated for the Calar Alto run 
by comparing a galaxy also observed in another calibrated 
Calar Alto run. 
For more details about the photometric calibration we refer to 
\cite{pohlen2001}.
%
\section{Extraction of the Disk Parameters}
\label{model}
\subsection{3-Dimensional Disk Model}
We have developed a semi-automatic recipe to fit true 3-D single-component 
luminosity distributions to the 2-D data of edge-on galaxies and determine 
the galaxy parameters, such as scalelength and scaleheight in a physically 
meaningful way.
Our method is described in detail in \cite{pohlen2000} and \cite{pohlen2001} 
and is only briefly recalled here. 
The disk model is based upon the fundamental work of 
\cite{vdk81}. 
They tried to find a fitting function for the three-dimensional 
light distribution in disks of edge-on galaxies using the 
empirically determined exponential radial gradient, $I\!\propto\!\exp(R)$,
and adding a description for the vertical distribution, $f(z)$, of the stars. 
The luminosity density distribution $\hat{L}(R,z)$ can be written as:
\begin{equation}
\nonumber
\hat{L}(R,z) = \hat{L}_0 \ \exp{\left(-\frac{R}{h}\right)} \ f_n(z,z_0) \ {\rm H
}(R_{\rm co}-R)
\nonumber
\label{hatl}
\end{equation}
$\hat{L}$ being the luminosity density in units of [$L_{\sun}$ pc$^{-3}$],
$\hat{L}_0$ the central luminosity density, $R$ and $z$ are the radial resp. 
vertical axes in cylinder coordinates, $h$ is the radial scalelength and 
$z_0$ the scaleheight, and $n$ the index of the vertical distribution 
function. H$(x_0-x)$ is the Heaviside function and \rco is the 
cut-off radius characterising the observed outer radial truncations 
\cite[][and references therein]{vdk79,pohlen2002}. 
To limit the choice of parameters we restrict our models to the 
three main density laws for the z-distribution ($\exp$, sech, and 
sech$^2$) following \cite{vdk88}. Due to the choice of our normalised 
isothermal case $z_0$ is equal to $2 h_z$, where $h_z$ is the usual 
exponential vertical scale height preferred by many authors: 
\begin{eqnarray}
\nonumber
f_1(z) &=& 4\ \exp{\left(-2\ \frac{\mid z \mid}{z_0}\right)} \\
\nonumber
f_2(z) &=& 2\ {\rm sech} \left(\frac{2 z}{z_0} \right) \\
\nonumber
f_3(z) &=& {\rm sech}^2{\left(\frac{z}{z_0}\right)}
\end{eqnarray}
For any details about the numerical realisation we refer to \cite{pohlen2000}.
Therefore six free parameters ($i,\ n,\ \hat{L}_0,\ R_{\rm co},\ h,\ z_0$) 
fit the observed surface intensity on the CCD chip to the model.
This model assumes that the vertical distribution is independent of 
position along the major axis as is known to be true in good approximation 
\citep[\cf e.g.][]{vdk81,shaw1990}. The increase of scaleheight with 
galactocentric distance as reported by \cite{degrijs1997} can be 
described as a combination of two disks each with constant scaleheight 
but different scalelength (\cf Sect.~\ref{tdds}). Any deviation from 
constant scaleheights should be visible in the vertical profiles overplotted
by the models, which is not the case (\cf Appendix \ref{app}). The apparent
small vertical shift of some models in the lowest plotted vertical profile 
is due to the change in radial scalelength, as described in Sect.~\ref{breaks}.
The possible influence on the six free parameters of the dust distribution,
which was neglected during the fit, is estimated in \cite{pohlen2000}. 
There we have shown that 
even for a worst case scenario (large optical depth $\tau_R$ and a radially 
and vertically fairly extended dust lane) our model is able to reproduce 
the input parameters with an error of a typical 20\%. We expect this 
effect to be even less significant for the present sample of lenticular, 
dust-depleted galaxies.  
Deriving individual errors on all parameters is a complex task 
since the main source of error is not the numerical fitting procedure 
($<\!1$\%) but the systematic uncertainties in the process of fitting 
a rather simple, empirical model to real galaxies. An estimation of 
the errors is given 
in \cite{pohlen2000} by changing the applied boundaries of the region used 
to fit the data to the model \cite[cf. also][]{pohlen2001}. 
They found differences in $h$ and $z_0$ of about $15\%$ and $\hat{L}_0$ 
varied about a factor of two.       
This is in the same range found by \cite{knapen1991} when they compared 
published values of scalelength measurements from different studies.    
\subsection{Thin/thick disks}
\label{tdds}
\cite{pohlen2001} noted that the \s0 galaxies in his sample are not 
well described by any combination of a single disk and another spheroidal 
component, such as a de Vaucouleurs $R^{1/4}$ bulge model 
\cite[]{devaucouleurs1948}.
All \s0 galaxies reveal a typical, continuous change of slope when one 
compares the major axis with parallel profiles above/below the plane. 
The profiles significantly flatten towards cuts higher above the midplane.
According to the single component model $\hat{L}(R,z)$ all slopes 
should be nearly parallel.  
We infer from the images (\cf Appendix \ref{app}) that all \s0 galaxies 
show a kind of smooth outer envelope or highly flattened spheroidal component. 
This deviation from a normal shape cannot be explained 
by a bulge component. Any large $R^{1/4}$ bulge would have to be 
apparent on the major axis which is certainly dominated by an 
exponential (disk) component. 
It is worth mentioning that in the sample of \cite{pohlen2001} there are 
four galaxies, NGC\,3390 ($T\eq3$), NGC\,3717 ($T\eq3$), 
NGC\,4696C ($T\eq3.4$), and NGC\,6504 ($T\eq2$)\footnote[1]{The classification 
is done by \cite{luett2000a}. It is consistent with that of 
\cite{vandokkum1994}, but significantly different from that of the UGC 
catalogue \cite[]{ugc}.} that showed 
a similar behaviour but were classified as late-type galaxies.
The outer component could be described as a `thick disk' according to 
\cite{bursteinIII} with a flatter slope than the inner thin disk, 
equivalent to a larger scalelength. However, the main characteristic
of a thick disk is that the observed vertical profiles depart from 
the simple exponential model. Compared to the model, the outer parts are 
systematically brighter with increasing distance to the major axis. 
As mentioned by \cite{degrijs1997}, 
the measured increase of the scaleheight with radius 
(`flaring') for early-type, edge-on disk galaxies can be understood if 
these galaxies have both thick disk scalelengths and scaleheights 
larger than for the dominant old disk. 
\subsection{Fitting Method}
\label{method}
To apply our single component fitting method described in the 
previous section we have to assume that there is a region in the galaxy 
dominated by only one of the components. If one uses exponential profiles it 
is obvious from the vertical and radial cuts that there are well separated 
vertical ranges (around the major axis and the outermost profiles) in which 
the light of one of the two disks dominates.
Fitting thin and thick disk component simultaneously would be the 
desired approach. However, our single component model 
has already six free parameters. As shown in \cite{pohlen2001}, 
fitting this model to observed data is a non-trivial task. The 
problem is the application of an idealized model itself, which 
obviously will not totally accurately describe the measured 
two-dimensional light distribution. This technique requires
continuous human supervision to control the influence and quality 
of each individual parameter. This would most probably be 
even more difficult for a parameter set twice as large. 
We decided therefore to use an iterative fitting routine starting from 
outside-in since the thick disk clearly dominates the outer parts. 
The first step is to determine an initial estimate for the outer disk by 
restricting the region to be fitted and using our single component 
model. The next step is subtracting the derived full thick disk model 
from the original image and fit an inner disk to the residual by 
restricting the fitting range to the inner parts.
Then we start from the beginning by subtracting this inner disk model
from the original image and fit again the outer disk, now to this 
residual image.
The initially pre-defined fitting regions for the thin and thick 
disk are sometimes adapted slightly after subtracting one of the 
components and before starting the second fitting round. 
As it turned out this process is remarkably stable for most of the 
galaxies. After one iteration the disk parameters are already 
the same within the range of one single model. 
The reason for this is the domination of the thick disk at 
large vertical z-heights ($\gtsim3.5\zn\eq7h_z^{\rm n}$) measured here
with high S/N, and the large radial range used to fit the thick disk 
component.
To restrict the number of free parameters we decided 
to prefix the inclination during the fitting process for this sample. 
For most of the galaxies the symmetric shape of the bulge and disk 
component indicates an exactly edge-on orientation. 
Only one of the \s0 galaxies (NGC\,3957) exhibits a dust-lane 
making it possible to estimate the inclination at $i\eq88\deg\pm1\deg$ 
following the method by \cite{BD1994}.   
The different density laws for the vertical distribution are similar 
for large z and only differ around the mid-plane of the luminosity 
distribution \citep[\cf Fig.4 in][]{degrijsetal1997}.
In \cite{pohlen2000} the actual choice of one fitting function is done
individually for each galaxy depending on the measured profiles.
However, near the plane of the galaxy the contribution of a thick disk 
is much smaller than that of to the thin disk component.
First tests with a free choice of the fitting function for the 
vertical density distribution ($\exp$, ${\rm sech}$, or ${\rm sech}^2$)
showed in the case of ESO\,311-012 that the iterative fit became unstable 
for this reason.
In the subsequent modelling we have chosen the intermediate fitting 
function $f_2(z)\!\propto\! {\rm sech}(2 z/z_0)$ in all cases for 
the thin and thick disk component. 
There was only one galaxy, NGC\,2310, for which the ${\rm sech}$ function 
did not yield a satisfying convergency of the iterative fitting. However, 
switching to ${\rm sech}^2$ significantly improved the fit.   
Finally, only three free parameters are left for each disk 
($\hat{L}_0,\ h,\ z_0$). 
We want to point out that these decompositions are thus model dependent. 
An isothermal thick disk contributes less light to the thin/thick disk 
combination than to an exponential one and the shape of the thin 
disk will also be different. 
In addition, vertical scaleheights $z_0$ obtained with an $\exp$ 
model are systematically larger than in a ${\rm sech}$ model
and again larger than in a ${\rm sech}^2$ fit. Depending on 
the vertical boundaries chosen for the fitting these differences 
could be more than $30\%$. 
The choice for the vertical distribution influences also the 
best-fit scale parameter ratio $\zkdzn$. \cite{shaw1989} used 
all combinations of either ${\rm sech}^2$ or $\exp$ models for 
NGC\,4565 and derived $\zkdzn$ in the range of 
$4.3-5.4$.
In contrast with some of the previous 1-D-only fitting methods we do 
not simply use the deviation from a simple exponential on the minor axis,
or parallel profiles; to measure thick disks we use the full 
2-dimensional information and are therefore able to fit the radial 
scalelength of the thick disks.
\subsection{Breaks in Radial Profiles}
\label{breaks}
One of the main difficulties while properly fitting our model to 
the \s0 galaxies is their outer disk structure. This will be 
discussed in detail in \cite{pohlen2004} and only briefly addressed here. 
As one can see from the figures in Appendix \ref{app} there are clear 
breaks in the outer parts of the radial profiles. These are 
similar to the truncations in more later-type galaxies 
\cite[cf.][]{pohlen2002,kregel2002,degrijs2001,pohlen2000a}.
For all galaxies (except ESO\,311-012) a similar break, slightly 
less pronounced, is also seen in other radial cuts parallel to the major 
axis. 
Our model, according to Eq. (\ref{hatl}), however, describes 
only an infinitely sharp truncation \rco. 
As shown in \cite{pohlen2001} this implies a tight coupling 
between the radial scalelength $h$ and the cut-off radius \rco,
when using our model fitting data with the observed breaks 
in the profiles.
In addition, the sharply truncated model exhibits an intrinsic bending 
of the profile towards the outer parts \cite[cf.][]{pohlen2001}. This 
complicates the visual quality control compared to the more flat 
infinite exponential model without any truncation.
Therefore we decided to use the infinite exponential model, 
realised within the same fitting program by fixing \rco to 
ten times the radial scalelength.
For our thick/thin fitting we restricted the fitting region 
to points inside the observed break radius. 
However, fitting an infinite exponential model to the 
intrinsic two-slope profiles is also affected by systematic 
errors \cite[cf.][]{pohlen2001,pohlen2004}. 
Depending on the ratio of the inner, shallow slope (\hin) up to the 
break, to the steeper, outer slope (\hout) beyond the break 
radius the determined scalelength will be systematically too small.  
Assuming a mean ratio of $\hindout\eq 4.4\!\pm\!1.7$ from a large, 
edge-on sample by \cite{pohlen2001} makes it possible to quantify the expected 
offset. The best fitting scalelength \hin will be about 26\% larger compared 
to the intrinsic radial scalelength \cite[cf.][]{pohlen2004}.  
Although the exact value of the scalelength is thus model dependent,
neither $z_0$ nor $\mu_0$ are influenced by this problem. 
The derived scaleheights are independent of the scalelength and 
therefore robust, but of course still depending on the chosen density 
law ($\exp$, sech, or sech$^2$) for the z-distribution. 
%
%
%
\section{Results}
\label{Results}
For six (75\%) out of the eight \s0 galaxies we were able to derive 
consistent thick/thin disk solutions (\cf\sec\ref{range}).
The resulting best fit parameters are listed in Table \ref{thickres} and 
Table \ref{thinres} for the thick and thin disk, respectively. 
\begin{table*}
\begin{center}
\begin{tabular}{lccc crc c crcc}
\hline
\rule[+0.4cm]{0mm}{0.0cm}
Galaxy
&Dist.
&Band
&\muk 
&\multicolumn{3}{c}{\zk}
&
&\multicolumn{4}{c}{\hk} \\
\cline{5-7}\cline{9-12}  \\[-0.15cm]
&\raisebox{0.0ex}{[Mpc]}
&\raisebox{0.0ex}{[$\lambda$]}
&\raisebox{0.0ex}{[\magsqarcsec]}
&\raisebox{0.0ex}{[\,\arcsec]}
&\raisebox{0.0ex}{[kpc]} 
&\raisebox{0.0ex}{[\zn]}
&
&\raisebox{0.0ex}{[\,\arcsec]}
&\raisebox{0.0ex}{[kpc]} 
&\raisebox{0.0ex}{[\hn]}
&\raisebox{0.0ex}{[\zk]} \\
\rule[-3mm]{0mm}{5mm}{\scriptsize{\raisebox{-0.7ex}{\it (1)}}} 
&{\scriptsize{\raisebox{-0.7ex}{\it (2)}}}
&{\scriptsize{\raisebox{-0.7ex}{\it (3)}}}
&{\scriptsize{\raisebox{-0.7ex}{\it (4)}}}
&\multicolumn{3}{c}{\scriptsize{\raisebox{-0.7ex}{\it (5)}}}
&
&\multicolumn{4}{c}{\scriptsize{\raisebox{-0.7ex}{\it (6)}}} \\
\hline \hline 
&&&&&&&&&&& \\[-0.3cm]
NGC\,2310\footnotemark[1] 
            &13.1&R&22.0&19.8 & 1.3& 2.4&& 62.8 &4.0$\,\;$& 2.6&3.2 \\
E\,311-012  &12.4&V&22.3&18.8 & 1.1& 2.6&& 41.5 &2.5$\,\;$& 1.9&2.2 \\
NGC\,3564   &36.7&V&21.9&22.0 & 3.9& 5.3&& 20.6 &3.7$\,\;$& 1.8&0.9 \\
NGC\,3957   &22.3&R&22.2&20.8 & 2.3& 3.4&& 33.9 &3.7$\,\;$& 1.6&1.6 \\
NGC\,4179   &17.7&V&22.9&37.3 & 3.2& 3.1&& 51.0 &4.4$\,\;$& 1.9&1.4 \\
NGC\,5047   &87.4&V&22.9&19.7 & 8.3& 3.4&& 39.4 &16.7$\,\;$& 1.7&2.0 \\
\hline
 mean         &    & &    &    &    &3.4&&     &   &1.9& 1.9\\
\hline
\hline 
\end{tabular}
\caption{Results for the thick disk: {\scriptsize{\it (1)}} Galaxy, 
{\scriptsize{\it (2)}} distance, {\scriptsize{\it (3)}} filter,
{\scriptsize{\it (4)}} central surface brightness (uncorrected for inclination) of the 
thick disk model \muk, {\scriptsize{\it (5)}} vertical scaleheight \zk in 
arcsec, parsec, and in units of the thin disk vertical scaleheight \zn, and 
the {\scriptsize{\it (6)}} radial scalelength \hk in arcsec, kpc, in units 
of the thin disk radial scalelength \hn, and in units 
of the thick disk vertical scaleheight \zk \hspace{0.5cm} $^{\rm a}$ (used sech$^2$ instead of sech)
\label{thickres} 
}
\end{center}
\end{table*}
\begin{table*}
\begin{center}
\begin{tabular}{lcc c r@{.}lc c ccc}
\hline
\rule[+0.4cm]{0mm}{0.0cm}
Galaxy
&Dist.
&Band
&\mun
&\multicolumn{3}{c}{\zn}
&
&\multicolumn{3}{c}{\hn} \\
\cline{5-7}\cline{9-11} \\[-0.15cm]
&\raisebox{0.0ex}{[Mpc]}
&\raisebox{0.0ex}{[$\lambda$]} 
&\raisebox{0.0ex}{[\magsqarcsec]}
&\multicolumn{2}{c}{\raisebox{0.0ex}{[\,\arcsec]}}
&\raisebox{0.0ex}{[kpc]} 
&
&\raisebox{0.0ex}{[\,\arcsec]}
&\raisebox{0.0ex}{[kpc]} 
&\raisebox{0.0ex}{[\zn]} \\
\rule[-3mm]{0mm}{5mm}{\scriptsize{\raisebox{-0.7ex}{\it (1)}}} 
&{\scriptsize{\raisebox{-0.7ex}{\it (2)}}}
&{\scriptsize{\raisebox{-0.7ex}{\it (3)}}}
&{\scriptsize{\raisebox{-0.7ex}{\it (4)}}}
&\multicolumn{3}{c}{\scriptsize{\raisebox{-0.7ex}{\it (5)}}}
&
&\multicolumn{3}{c}{\scriptsize{\raisebox{-0.7ex}{\it (6)}}} \\
\hline \hline 
&&&\multicolumn{2}{}&&&& \\[-0.3cm]
NGC\,2310     &13.1&R& 19.6& 8&1 &0.5&& 24.0&  1.5 &  3.0 \\
ESO\,311-012  &12.4&V& 20.5& 7&2 &0.4&& 21.4&  1.3 &  3.0 \\
NGC\,3564     &36.7&V& 19.5& 4&2 &0.7&& 11.3&  2.0 &  2.7 \\
NGC\,3957     &22.3&R& 19.2& 6&0 &0.7&& 21.6&  2.3 &  3.6 \\
NGC\,4179     &17.7&V& 20.2&12&1 &1.0&& 26.2&  2.3 &  2.2 \\
NGC\,5047     &87.4&V& 20.7& 5&8 &2.5&& 22.6&  9.6 &  3.9 \\
\hline 
 mean             &    & &     &  & &   &&     &     &  3.1\\
\hline
\hline 
\end{tabular}
\caption{Results for the thin disk: {\scriptsize{\it (1)}} Galaxy, 
{\scriptsize{\it (2)}} distance, {\scriptsize{\it (3)}} filter,
{\scriptsize{\it (4)}} central surface brightness (uncorrected for inclination) of the 
thin disk model \mun, {\scriptsize{\it (5)}} vertical scaleheight \zn in 
arcsec, and parsec, and the {\scriptsize{\it (6)}} radial scalelength \hn 
in arcsec, kpc, and in units of the thin disk vertical scaleheight \zn 
\label{thinres} }
\end{center}
\end{table*}
The resulting radial and vertical profiles overplotted by our best 
fitting model are shown in Appendix \ref{app}.
The derived thin disk scalelength and scaleheight values are consistent
with those found in studies by \cite{kregel2002} and \cite{pohlen2001}, 
containing mostly galaxies of Hubble type later than Sa. 
The mean ratio of thick to thin scaleheight for the five galaxies 
with the same vertical model combinations (sech$+$sech) for the thick 
and thin disk is $\zkdzn = 3.6 \pm 1.0$ and for the scalelength 
$\hkdhn = 1.8 \pm 0.1$.
Including also NGC\,2310 we find $\zkdzn = 3.4 \pm 1.0$ and 
$\hkdhn = 1.9 \pm 0.4$. 
We derive central surface brightnesses (uncorrected for inclination) 
in the range 
 of $22.0\ltsim\muk\ltsim22.2$\,R-\magsqarcsec 
and $21.9\ltsim\muk\ltsim22.9$\,V-\magsqarcsec for our thick disks 
compared to $19.2\ltsim\mun\ltsim19.6$\,R-\magsqarcsec 
and $19.5\ltsim\mun\ltsim20.7$\,V-\magsqarcsec for the thin disks. 
This implies that the contribution of the thick disk to 
the central surface brightness is about 10\% of that of the thin disk. 
The thick disk central luminosity density $\hat{L}_0^{\rm k}$ ranges 
between 3.5\% and about 10\% (mean: 5.6\%) of the thin disk value. 
The ratio of the total luminosities of the thick and thin disk is between 
about one third and one.
The profile on the minor axis in addition to radial cuts 
high above the major axis (\cf Appendix \ref{app}) reveals that for 
four (NGC\,2310, NGC\,3564, NGC\,4179, NGC\,5047) out of the 
six fitted galaxies there is no significant 
bulge component visible at large vertical height above the disk. 
Even the apparently bulge dominated galaxy NGC\,3564 is well
described by a thick disk component. The remaining bulge light 
after subtraction of the thin/thick disk combination is 
comparable to that of the other galaxies. 
This implies that all bulges of our \s0 galaxies could not
be well described by a traditional de Vaucouleurs $R^{1/4}$ bulge. 
In addition, any bulge component would be too flat to account for 
all the light high above the disk at large galactocentric radii.
%
%
\subsection{Fitting regions}
\label{range}
\begin{table}
\begin{center}
\begin{tabular}{l r@{-}l r@{-}l c r@{-}l r@{-}l}
\hline
\rule[+0.4cm]{0mm}{0.0cm}
Galaxy
&\multicolumn{4}{c}{vertical region $z$}
&
&\multicolumn{4}{c}{radial region $R$}  \\
&\multicolumn{2}{c}{thin}
&\multicolumn{2}{c}{thick}
&
&\multicolumn{2}{c}{thin}
&\multicolumn{2}{c}{thick} \\
\cline{2-5}\cline{7-10} \\[-0.15cm]
&\multicolumn{4}{c}{\raisebox{0.0ex}{[\,\arcsec]}}
&
&\multicolumn{4}{c}{\raisebox{0.0ex}{[\,\arcsec]}} \\
\rule[-3mm]{0mm}{5mm}{\scriptsize{\raisebox{-0.7ex}{\it (1)}}} 
&\multicolumn{4}{c}{\scriptsize{\raisebox{-0.7ex}{\it (2)}}}
&
&\multicolumn{4}{c}{\scriptsize{\raisebox{-0.7ex}{\it (3)}}} \\
\hline \hline
&\multicolumn{4}{c}{} &  \multicolumn{4}{c}{} \\[-0.3cm]
NGC\,2310      &\hspace{0.3cm}0   &12   &20   &35 &   &32   &75  &  0 &  79 \\
ESO\,311-012   &0   & 8   & 8   &31 &   &29   &88  & 37 & 117\\
NGC\,3564      &0   & 4   &22   &32 &   &20   &32  &  0 &  52\\
NGC\,3957      &0   & 4   &16   &28 &   &36   &56  & 24 &  64\\
NGC\,4179      &0   &12   &31   &55 &   &43   &69  &  0 &  88\\
NGC\,5047      &0   & 6   &18   &25 &   &35   &47  &  0 &  55\\
\hline
\end{tabular}
\caption{Radial and vertical fitting regions for the thick/thin disk 
components with {\scriptsize{\it (1)}} galaxy name, {\scriptsize{\it (2)}} 
beginning and end of thick/thin disk vertical region, and 
{\scriptsize{\it (3)}} beginning and end of thick/thin disk 
radial region \label{regions}}
\end{center}
\end{table}
One of the important constraints for fitting empirical, surface photometric 
models to observed data is the definition of the actual regions which 
are marked to characterise the individual model components. 
Therefore we list in Table \ref{regions} for each galaxy the radial and 
vertical ranges where our thick/thin disk fit was applied.  
In all cases we restricted the two fitting regions to be distinct 
from each other and outside the dust lane (in the case of NGC\,3957) 
or any bar/ring like feature visible in the radial profile.  
This is obvious for the first thick disk estimation.
However, in principle one could extend the fitting range towards the 
inner/outer parts for the following thick/thin disk iterations, since 
the other component in each case has already been subtracted.
In addition, we masked by hand bright stars and background galaxies, 
forcing the program to ignore these regions. 
%
\subsection{Comparison with Literature}
\begin{table*}
\begin{center}
\begin{tabular}{l l c c c  c  c  l}
\hline
\rule[+0.4cm]{0mm}{0.0cm}
Galaxy
&Type
&Model
&\zkdzn
&\hkdhn 
&\muk [\magsqarcsec] 
&\mun [\magsqarcsec]
&Reference  \\[+0.1cm]
&
&
&
&
&{\scriptsize \raisebox{-0.9ex}{$\stackrel{\rm R}{\rm V}\:$}}-band 
&{\scriptsize \raisebox{-0.9ex}{$\stackrel{\rm R}{\rm V}\:$}}-band
& \\
\rule[-3mm]{0mm}{5mm}{\scriptsize{\raisebox{-0.7ex}{\it (1)}}} 
&{\scriptsize{\raisebox{-0.7ex}{\it (2)}}}
&{\scriptsize{\raisebox{-0.7ex}{\it (2)}}}
&{\scriptsize{\raisebox{-0.7ex}{\it (3)}}}
&{\scriptsize{\raisebox{-0.7ex}{\it (4)}}}
&{\scriptsize{\raisebox{-0.7ex}{\it (5)}}}
&{\scriptsize{\raisebox{-0.7ex}{\it (6)}}}
&{\scriptsize{\raisebox{-0.7ex}{\it (7)}}} \\
\hline \hline 
&&&&&& \\[-0.3cm]
5 galaxies\footnotemark  
&S0  
&sech$+$sech
&$2.6\leftrightarrow5.3$ 
&$1.7\leftrightarrow1.9$ 
& \raisebox{-0.7ex}{$\stackrel{22.2}{\scriptstyle 21.9\leftrightarrow22.9}$}
& \raisebox{-0.7ex}{$\stackrel{19.2}{\scriptstyle 19.5\leftrightarrow20.7}$ }
& this study \\[+0.1cm]
NGC 4710& S0 &sech$+$exp&3.2&&&& DE96 \\[+0.1cm]
NGC 4762& S0 &sech$+$exp&4.6&&&& DE96 \\[+0.1cm]
5 galaxies  
&S0  
&$-$
&1.8$\leftrightarrow$4.6 
&            
&        
& 
& DE97 \\[+0.1cm]
NGC\,6504   
&Sab 
&$\exp+\exp$
&4.0         
&            
&\raisebox{-0.7ex}{$\stackrel{23.8}{\scriptstyle - }$ }     
&\raisebox{-0.7ex}{$\stackrel{18.2}{\scriptstyle - }$ }
&VA94 \\[+0.1cm]
NGC\,891   
&Sb  
&$\exp+\exp$
&2.3$\leftrightarrow$6.3     
&
&
&
& MO97\\[+0.1cm]
NGC\,891\footnotemark   
&Sb  
&$R^{1/4}+$sech$^2+\exp$
&3.0
&
&
&
& VA84\\[+0.1cm]
NGC\,4565\footnotemark   
&Sb  
&sech$^2+$sech$^2+$halo
&2.2         
&1.4
&     
&       
& WU02\\[+0.1cm]
NGC\,4565 
&Sb  
&sech$^2+$sech$^2$
&4.6         
&1.4         
&\raisebox{-0.7ex}{$\stackrel{23.7}{\scriptstyle - }$  }    
&\raisebox{-0.7ex}{$\stackrel{19.6}{\scriptstyle - }$  }      
& SH89 \\[+0.1cm]
NGC\,4565 
&Sb  
&$\exp+\exp$
&5.4         
&1.1         
&\raisebox{-0.7ex}{$\stackrel{23.0}{\scriptstyle - }$  }   
&\raisebox{-0.7ex}{$\stackrel{18.8}{\scriptstyle - }$  }     
& SH89 \\[+0.1cm]
MW\footnotemark (optical)&Sbc 
&$\exp+\exp$
&3.0         &1.3         &         &       & LA03\\[+0.1cm]
MW\footnotemark[4] (NIR)   &Sbc 
&$\exp+\exp$
&3.3         &1.3         &         &       & OJ01\\[+0.1cm]
ESO\,342-017
&Scd 
&$\exp+\exp$
&2.5    
&\gtsim1.0        
&\raisebox{-0.7ex}{$\stackrel{22.1}{\scriptstyle22.5 }$ }
&\raisebox{-0.7ex}{$\stackrel{20.0}{\scriptstyle20.5 }$ }
& NE02\\[+0.1cm]
IC\,5249  
&Sd  
&$\exp+\exp$
&3.0         
&0.6         
&   
&       
& AB99\\
\hline
\end{tabular}
\caption{
Comparison with literature:
 {\scriptsize{\it (1)}} Galaxy name 
{\scriptsize{\it (2)}} morphological type, {\scriptsize{\it (3)}} ratio 
of thick to thin disk scaleheight, {\scriptsize{\it (4)}} ratio 
of thick to thin disk scalelength, {\scriptsize{\it (5)}} central surface 
brightness of thick disk (if quoted), {\scriptsize{\it (6)}} central surface 
brightness of thin disk, 
{\scriptsize{\it (7)}} reference DE96:\cite{degrijs1996}, DE97:\cite{degrijs1997}, VA94:\cite{vandokkum1994}, WU02:\cite{wu2002}, MO97:\cite{morrison1997}, VA84:\cite{vdk84}, SH89:\cite{shaw1989}, LA03:\cite{larsen2003}, OJ01:\cite{ojha2001}, NE02:\cite{neeser2002}, AB99:\cite{abe1999} \label{comparison} \newline
{\footnotesize
{$^{\rm a}$} NGC\,2310 excluded (only sech$^2$ model for thick disk) 
\hspace{0.43cm} {$^{\rm c}$} three component model (thin disk, thick disk, and halo)   \newline
{$^{\rm b}$} three component model (bulge and thin plus thick disk) 
{$^{\rm d}$} starcount analysis 
}
}
\end{center}
\end{table*}
\cite{bursteinIII} only describes the `thick disk' component qualitatively
as being more diffuse than that of to the inner, dominating disk, and 
possessing 
a flattened shape. The important observation is that the scaleheight of a 
fitted exponential to the vertical profiles increases with radial 
distance $R$. 
He determines a flattening, defined as the intrinsic z-thickness  
compared to the diameter $a$, of $z/a\eq0.25 - 0.2$ at $\mu_{\rm B}\eq 25.0$ 
which is somewhere between thin disks ($z/a\ltsim0.1$) and E4 ellipticals 
($z/a\eq0.6$) for five edge-on \s0 galaxies.
We find values of 0.29 (NGC\,2310) or 0.32 (NGC\,5047) comparing 
minor to major axis diameter at the second outermost
contour (\cf\fig\ref{n2310},\ref{n5047}).
While comparing scale parameter ratios in the literature one has to 
keep in mind that they are often obtained with very different fitting 
methods and even different fitting functions for the vertical distribution
of thin and thick disk light. As discussed in Sect.~\ref{method} this 
already implies systematic differences of at least $20\%$. 
Surprisingly, there are no detailed parameter studies for thick disks 
of \s0 galaxies in the literature that provide both
scaleheight and scalelength ratios (\cf Sect.~\ref{introduction}). 
The only scaleheight ratios for \s0 galaxies (\cf Table \ref{comparison}) 
are given for two galaxies (NGC\,4710, NGC\,4762) in detail 
by \cite{degrijs1996} and for five galaxies with Hubble type 
$T\!\ltsim\!0.0$ without providing the individual values or applied 
fitting functions by \cite{degrijs1997}. 
They find ratios of $\zkdzn\eq1.8-4.6$, which falls well within the range
of values for our \s0 galaxies. \newline
\indent There are clearly more studies in the literature that provide
detailed parameters (scaleheights and often also scalelengths) for thick 
disks in galaxies of later morphological type. \newline
\indent \Citet{vandokkum1994} modeled excess light at large vertical 
distance one-dimensionally with a thin/thick disk combination
for the edge-on Sab galaxy NGC\,6504. 
They found that the ratio of scaleheights is roughly 4, slightly 
higher than the mean value for our \s0 sample. 
However, their central surface brightness for the thick disk 
is significantly lower ($\Delta \mu_0\eq1.6$\,mag) than our value 
for NGC\,3957. 
This effect is not related to the different fitting functions used
since our sech model should tend to result in systematically lower 
values for $\mu_0$ than in their $\exp$ model. \newline
\indent For the Milky Way, recent optical star-count measurements 
by \cite{larsen2003} yield thin disk values of $\zn\eq300$\,pc 
and $\hn\eq3.5$\,kpc, and for the thick disk $\zk\eq900$\,pc 
and $\hk\eq4.7$\,kpc. These are similar to the 
infrared 2MASS star counts by \cite{ojha2001} with $\zn\eq260$\,pc, 
$\hn\eq2.8$\,kpc, $\zk\eq860$\,pc, and $\hk\eq3.7$\,kpc, keeping in 
mind that the near-infrared surveys always derive systematically 
smaller scalelengths than in the optical.
The resulting ratio of thick to thin disk scaleheight is very similar 
to the mean value we find for our \s0 galaxies but their ratio of the 
scalelengths is also smaller.
\cite{du2003} present a list of thick disk local normalizations 
(relative to the solar neighbourhood) obtained in different studies
including their own measurements (\cf their Table 2).
The quoted values range from 2\% up to 13\% with a mean value 
of about 6.1\%, which agrees very well with the mean value of 5.6\% 
for the central luminosity density of our \s0 galaxies. \newline
\indent \cite{morrison1997} used one-dimensional vertical fits to derive the 
thick and thin disk scaleheights of NGC\,891, an edge-on Sb galaxy 
similar to our Milky Way. They quote a large range of possible 
$\zkdzn$ values which are consistent with our \s0 disks.  
Previous modeling of NGC\,891 by \cite{vdk84} yield a ratio of 
$\zkdzn\eq3.0$ for a three component (thin plus thick disk, and 
additional $R^{1/4}$-spheroid) fit which is also in the range of 
our \s0 galaxies. The slightly higher value of the thick 
disk central luminosity density ($17\%$ of the thin disk) 
compared to our maximum of $10\%$ could be explained by the 
additional contribution of the applied $R^{1/4}$-spheroid, especially 
high above/below the midplane.   
However, \cite{bahcall1985} already pointed out that in the case of 
NGC\,891 a combination of thin plus thick disk and a model with only
one disk and a $R^{1/4}$-spheroid both give a good description of
the data used by \cite{vdk84}. \newline
\indent For another prominent edge-on Sb galaxy, NGC\,4565, several 
attempts for a decomposition are available. More recently, \cite{wu2002} 
presented a thin/thick plus halo decomposition from a deep intermediate-band 
(6660\AA) image deriving values only slightly smaller than for 
our \s0 galaxies (\cf Table \ref{comparison}). 
However, the two component disk models  of NGC\,4565 by \cite{shaw1989} yield 
significantly higher values for the ratio $\zkdzn$.
The light \cite{wu2002} put down to a halo is here ascribed solely 
to a thick disk with a large scaleheight. There is a large variety 
of model combinations (with different components) possible 
\citep[cf.][]{shaw1989,naeslund1997}, which complicates a direct 
comparison. \newline
\indent NGC\,5907, another nearby edge-on Sc galaxy, is an even 
more complex case. It is not yet clear if the multiple 
detected extended light distribution is a thick disk or
a stellar halo component \cite[cf.][]{morrison1994}. 
\cite{zheng1999} even concluded that NGC\,5907 does not have 
a faint extended halo at all (but compare with discussion in 
\cite{neeser2002} for this issue).  \newline
\indent \cite{abe1999} obtained a deep optical image of the edge-on Sd 
galaxy IC\,5249. They detected additional light to that predicted 
by a single exponential disk and tried to fit a thick disk model.
The scaleheight ratio of their best thick disk is well within 
the range for our \s0 galaxies.
However, the scalelength of the thick disk is exceptionally smaller compared 
to the thin disk and their central luminosity density 
($\hat{L}_0^{\rm n}/\hat{L}_0^{\rm k}\eq7.4$) is noticeable larger 
than for our sample. \newline 
\indent Recently, \cite{neeser2002} reported for the first time 
the detection of a thick disk in a low surface brightness 
galaxy (ESO\,342-017 classified as Scd). They find for the scaleheight 
a ratio of $\zkdzn \eq 2.5 $ (close to our values for NGC\,2310 and 
ESO\,311-012) with a comparable or somewhat larger scalelength for 
the thick disk.  
The scaleheight ratio and unprojected central surface brightness
(\cf Table \ref{comparison}) are surprisingly similar to
the range of values found for our \s0 galaxies.
%
%
%
%
\section{Discussion} 
\label{Discussion}
\subsection{Thick disks: discrete or continuous?}
A key question is: {\sl How to describe a thick disk in 
general? } With reference to the proposed different formation 
scenarios described in  Sect.~\ref{introduction} we can assume that 
the disk component is either  
characterised as a superposition of two discrete and independent 
isothermal disk systems (as done here), or built from the contribution 
of multiple velocity-dispersion components \cite[e.g.][]{wielen1992,dove1993}.
These two approaches on how to treat a thick disk seem to be incompatible. 
Therefore any derived parameters for the two-disks model appear useless 
for the multi-component disk. 
In addition, the latter seems to be superior since it is consistent with the 
model of continuous disk heating leading to the well known observable 
age-velocity dispersion relation \cite[]{wielen1977}.
However, \cite{majewski1993} already states in his review that detailed studies
of the spatial and kinematical distribution of stars in the Milky Way do not 
make it possible to decide if the thick disk is a discrete component or a more 
continuous sequence of stellar populations. 
Recently, \cite{nissen2003} concluded that the latest studies \citep[e.g.][]{bensby2003} argue for the separate entity picture.
This confirms our result that 75\% of the chosen \s0 galaxies are well 
described with a distinct, two-component thin/thick disk system.
Note that these results exclude all thick disk
formation scenarios based solely on heating. 
Especially the elemental abundance trends found by \cite{bensby2003} 
favour a merger scenario where a satellite galaxy either merges with 
the parent galaxy or sheds significant amounts of its material to
form the thick disk as proposed by \cite{abadi2003}.    
However, as already noted by \cite{jacobi1994} it does not seem possible 
to distinguish between our simple model and more sophisticated ones.     
Therefore the very good description with the applied discrete 
two-component system for our galaxies is only a first step towards studying 
the luminosity distribution of external galaxies.
In addition, one has to keep in mind that our model only fits 
well in a restricted radial range (only out to $2.2$\hn$-3.9$\hn)
where most of the inner part is ``hidden'' by the 
central bulge, bar, or ring components. 
One key to the nature of the thick disk may lie in studying 
the outer parts where the breaks in the radial profiles of both 
thin and thick disk should provide an additional constraint. 
Any vertical colour gradients could provide additional 
information but the only available colour map for our 
sample \cite[NGC\,3957 in][]{pohlen2001} suffers from 
low S/N and is not conclusive. 
\subsection{Are thick disks of \s0s exceptional ?}
The fact that the range of thick disk parameters for all known \s0 
galaxies are not too different from those of 
late-type galaxies, even compared to a low surface brightness Scd galaxy, 
is especially surprising, since  at first glance one expects \s0 
galaxies to posses more prominent thick disks.  
However, in terms of the thick-to-thin-disk scaleheight ratio our values 
agree well with those derived for all other galaxies 
(\cf Table \ref{comparison}).
Even taking into account the central surface brightness or 
central luminosity density the comparison yields similar values. 
Does this point to a general formation process for thick disks
independent of bulge-to-disk ratio and Hubble type ? 
At this stage we are not able to answer this question. Although 
the numbers of galaxies used for the comparison are the same, all
literature values come from different sources, using sometimes 
very different methods and models to derive their parameters. In 
the case of NGC\,4565 it is obvious that depending on the model, 
one can find an even larger variety of $\zkdzn$-values than the 
full range for our six galaxies. 
In addition, as pointed out by \cite{knapen1991}, and again 
discussed in detail by \cite{pohlen2001}, 
fitting disk-model components to surface 
brightness data of edge-on galaxies is a delicate business, and 
comparing the results of different authors can be misleading. 
To overcome this problem one has to extend this survey of 
thick disk parameters also to late-type galaxies. To reduce 
the inevitable influence of their prominent dust lanes this has to 
be done in the near-infrared. After applying the same fitting method 
one is able to address the questions if thick disks are really common 
around all Hubble-types, and if their parameters are really similar,
as suggested here. This would entail a common formation scenario for 
thick disks independent of their normal evolution along the Hubble 
sequence.    
Of our galaxies with successful thick disk models there are only two 
for which there are rotational velocity measurements in the
literature. Therefore any correlation between mass and thick
disk parameters could not be derived. Consequently it is still 
unclear if the mass, as a galaxy characteristic, is related 
to the thick disk parameters. 
\subsection{Thick disk scalelengths}
We find significantly larger scalelength ratios ($\hkdhn\!=\!1.9$) for 
the thick disks of our \s0 galaxies than in the literature 
($\hkdhn\!\ltsim\!1.4$). 
Is this a possible distinction in thick disk parameters between late-type 
and our early-type galaxies ? 
Again, differences in fitting methods, especially for the scalelength in 
edge-on galaxies, could be responsible for this apparent 
disagreement.
In particular, fitting profiles with clear breaks using a single disk 
(with an infinitely sharp truncation or infinite exponential as done 
here) will entail systematic errors (\cf \sec\ref{breaks}), therefore 
this question could only be addressed unambiguously applying exactly the 
same method to late-type galaxies. 
However, although the exact number is uncertain the scalelength of the 
thick disk is without doubt systematically 
larger than that of the thin disk. Except in one case this is also true for 
all the literature values. Does this larger scalelength imply a different 
formation process or even restrict the available formation scenarios ?
At a first glance the different scalelengths contradict a dynamical 
heating scenario, since this should only alter the vertical distribution. 
However, assuming two distinct disk components with radially constant 
thickness the velocity dispersion scales with radius \cite[]{vdk86}.
It depends therefore on the exact way (radial distribution) the proposed 
mechanisms (\cf\sec\ref{introduction}) dispenses the energy in the 
vertical motions of the stars. 
In the case of heating the disk by satellite accretion, the N-body 
simulations of \cite{quinn1993} show that the scalelength of the 
disk is nearly unchanged (even slightly smaller than the 
initial scalelength) for the inner parts (out to $3h_{\rm ini}$).
Due to the migration of material outwards in radius, at larger radius 
(out to $\approx\!6h_{\rm ini}$) the final disk shows a second shallower 
component (\cf their Fig.~4). 
However, \cite{aguerri2001} find in their N-body simulation a global 
outward transport of disk material leading to a general increase of 
the disk scalelength of 10\% to  60\%.   
Although this increase is below our measured scalelength ratios, the latter 
seems to coincide in general with observations of larger thick disk 
scalelengths. 
One has to keep in mind that in the satellite accretion scenario a pre-existing
thin disk gets heated. The thin disk to be observed today must be
rebuilt out of gas remaining after the merger process.
It is not guaranteed that this thin disk will have the same 
scalelength as its predecessor since we do not have a unique 
explanation of how the disk scalelength is determined out of 
an initial (or new) gas distribution \cite[cf.][]{pohlen2000}.  
Looking from a slightly different angle, one can also try to apply 
the larger thick disk scalelength as a general argument against an 
internal heating scenario.
Assuming infinitely exponential disks, the larger scalelength 
implies a larger angular momentum content of the thick disk. Any valid 
heating mechanism must therefore add angular momentum, which is not the 
case for the internal heating scenarios. 
However, note that the disks are not at all infinitely exponential
but show clear breaks in their profiles (\cf\sec\ref{breaks}).
The disks exhibit the break at roughly similar radial radius 
(\eg\fig\ref{n5047}). In this sense the thick disk is truncated 
``earlier'' (in respect to its scalelength) than the thin disk.
In addition to a probably lower rotational velocity this could add up 
again to a similar angular momentum content for thick and thin disk,
ruling out the whole argument against the heating scenarios.   
The key point here is again the origin of these breaks 
in the radial profiles. 
\subsection{Residuals}
In addition to comparing thick disks across a large range of 
characteristic parameters of galactic disks, fitting and 
subtracting a combined disk yields valuable information on 
the structure and size of galactic bulges. 
As discussed in Appendix \ref{individual} the residual images 
highlight the deviation of the galaxy from the disk model. 
Structures only faintly indicated in the profiles become obvious.   
One important result related to this is that for our sample of 
\s0 galaxies none of the present bulges could be described as 
a traditional early-type $R^{1/4}$ bulge in agreement with 
\cite{balcells2003}.
However, a detailed structural analysis of the remaining central 
structures has to be done with great care. As discussed in 
\sec\ref{method}, choosing the vertical model is not a unique 
process and therefore any variation of $f(z)$ for the thick 
and thin disk could alter the shape of the disk profile in 
the inner bulge parts. Especially for galaxies where
the bulge component along the minor axis extends vertically 
above the disk (ESO\,311-012 and NGC\,3957) this could be 
even more unconstrained. 
%
%
%
%
%
\begin{acknowledgements}
We thank Reynier Peletier for stimulating discussions and acknowledge
useful suggestions by the referee, Richard de Grijs. 
Part of this work was supported by the German
\emph{Deut\-sche For\-schungs\-ge\-mein\-schaft, DFG}. 
This research has made use the Lyon/Meudon Extragalactic Database 
(LEDA, {\ttfamily http://leda.univ-lyon1.fr}) and the NASA/IPAC Extragalactic 
Database (NED) 
which is operated by the Jet Propulsion Laboratory, 
California Institute of Technology, under contract with the National 
Aeronautics and Space Administration. 
It also uses the Digitized Sky Survey (DSS) 
based on photographic data obtained using Oschin Schmidt Telescope on Palomar 
Mountain and The UK Schmidt Telescope and produced at the Space Telescope 
Science Institute. This research has made use of NASA's Astrophysics Data 
System Bibliographic Services.
\end{acknowledgements}

%
\appendix
\section{Comments on Individual Galaxies}
\label{individual}
\begin{figure}
\includegraphics[width=5.8cm,angle=270]{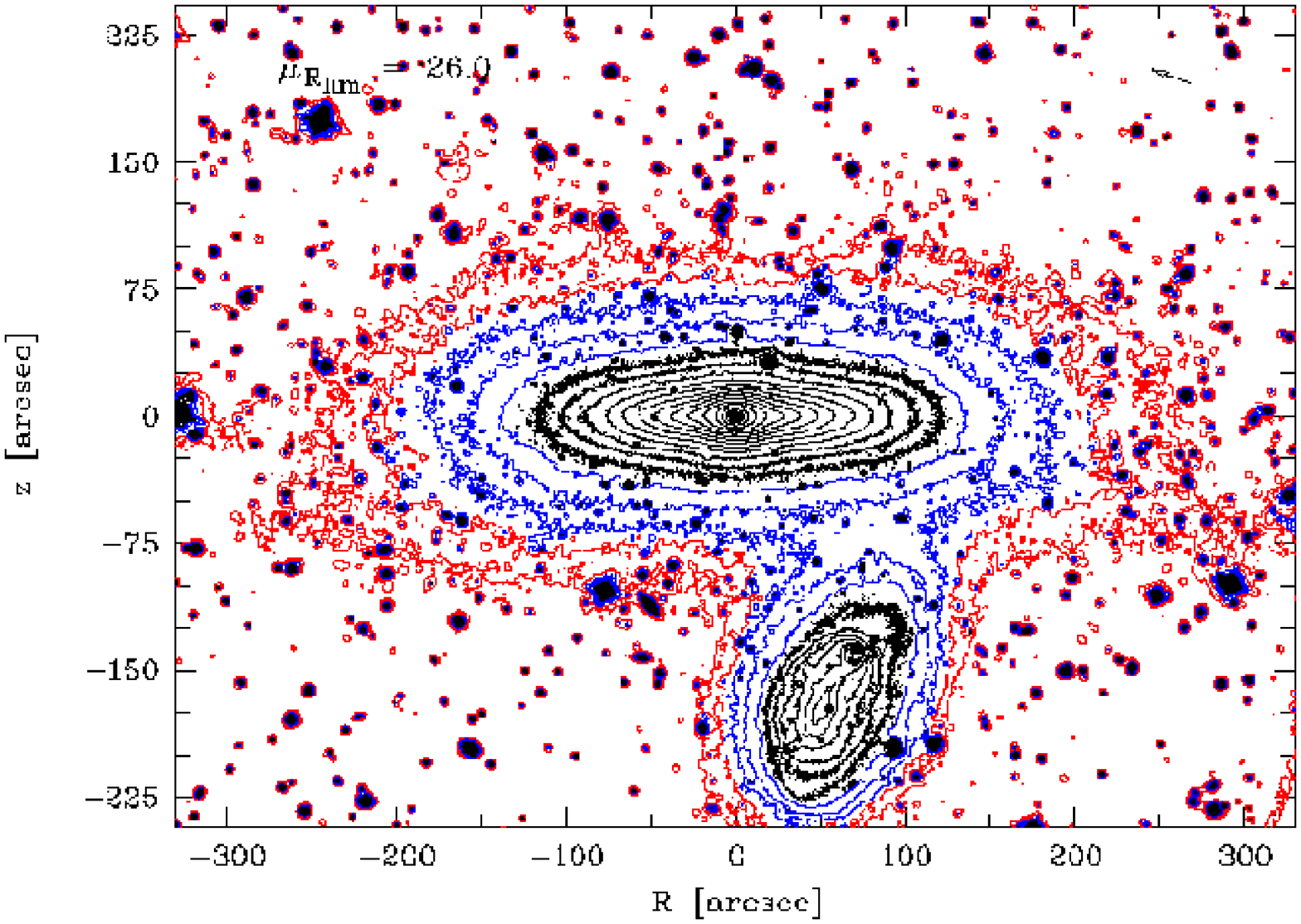}
\includegraphics[width=5.8cm,angle=270]{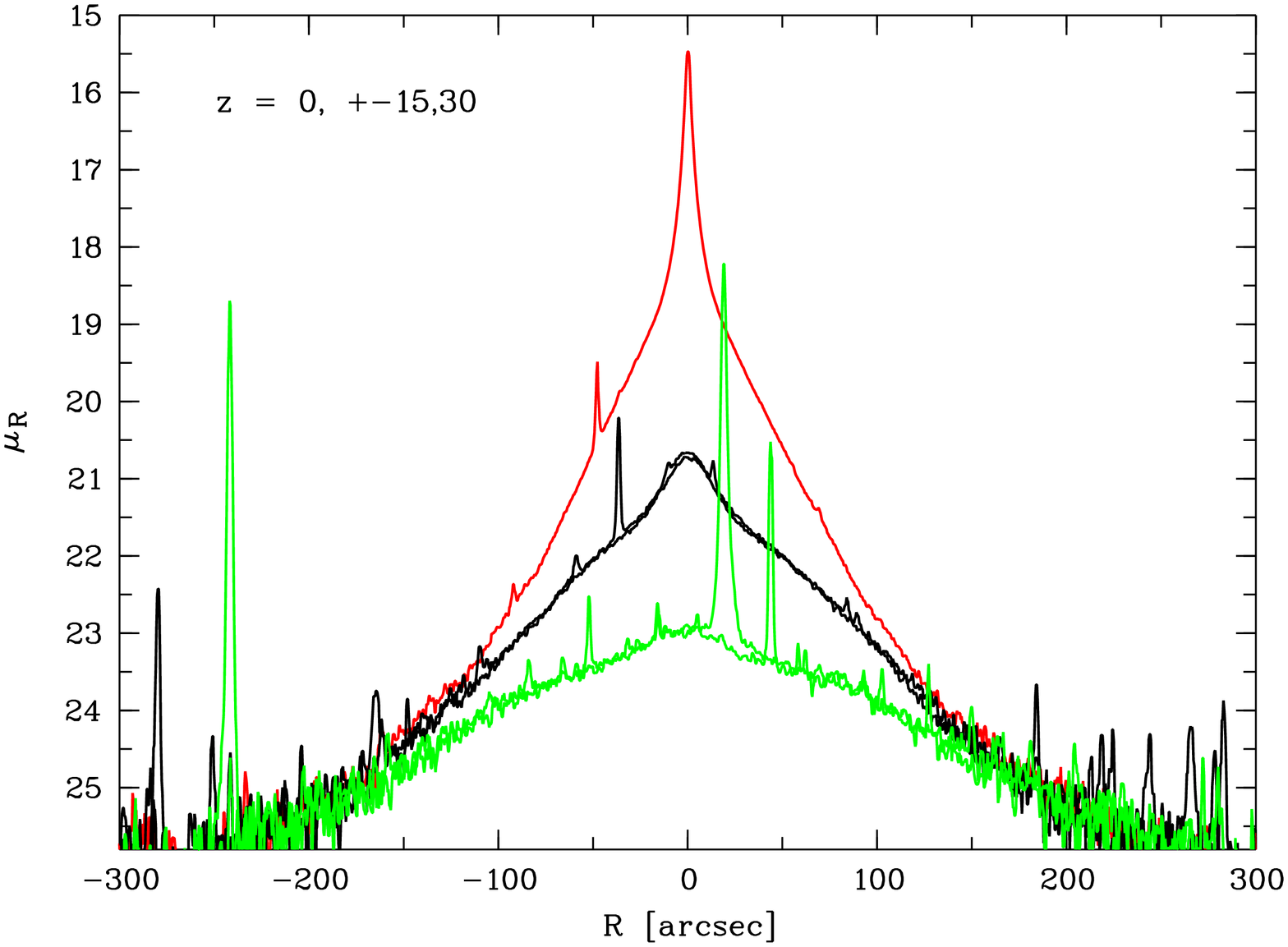}
\includegraphics[width=5.8cm,angle=270]{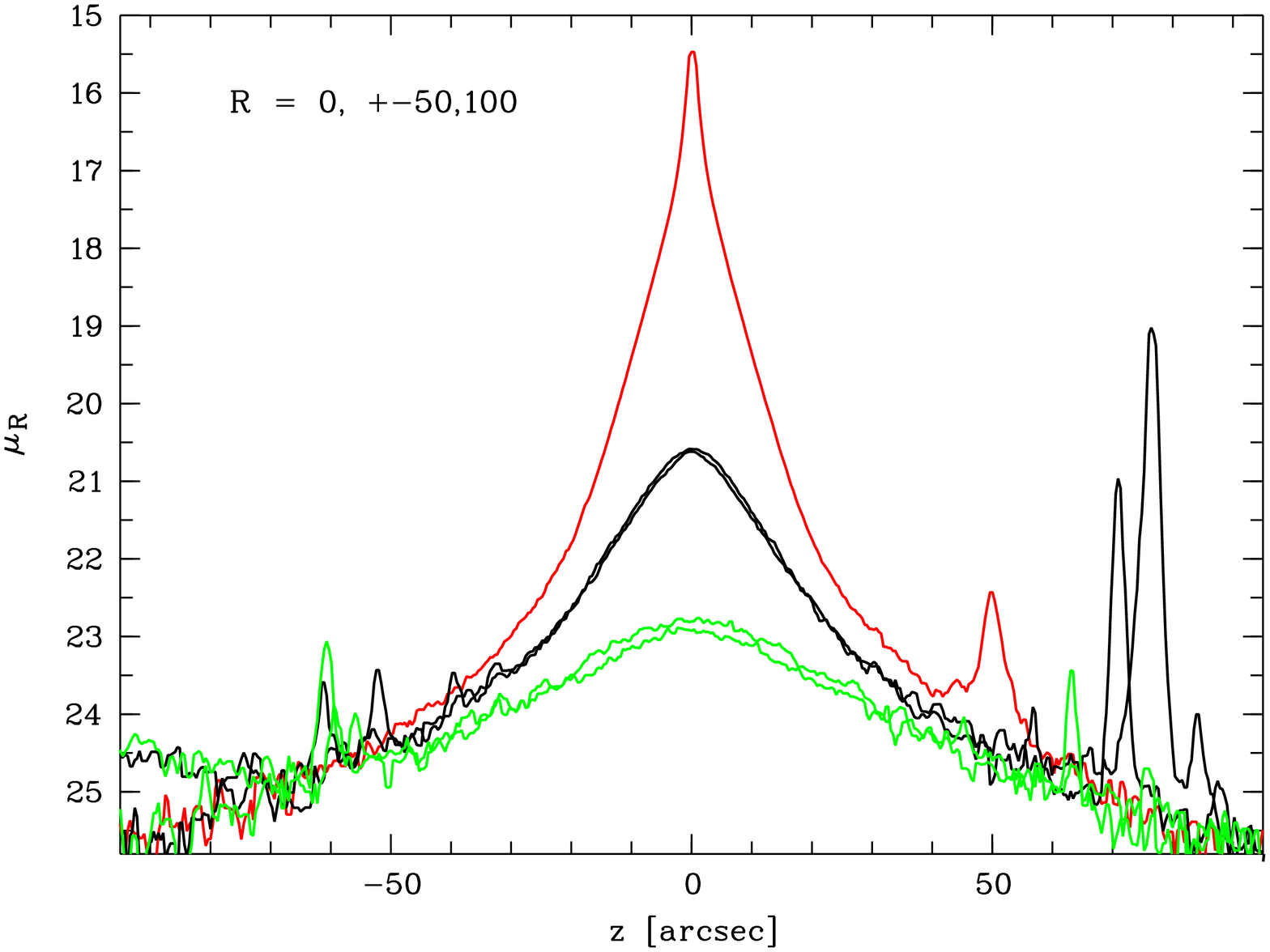}
\vspace*{5.85cm} 
\caption{NGC\,1596 R-band \label{n1596}}
\end{figure}

\begin{figure}
\newpage
\includegraphics[width=5.8cm,angle=270]{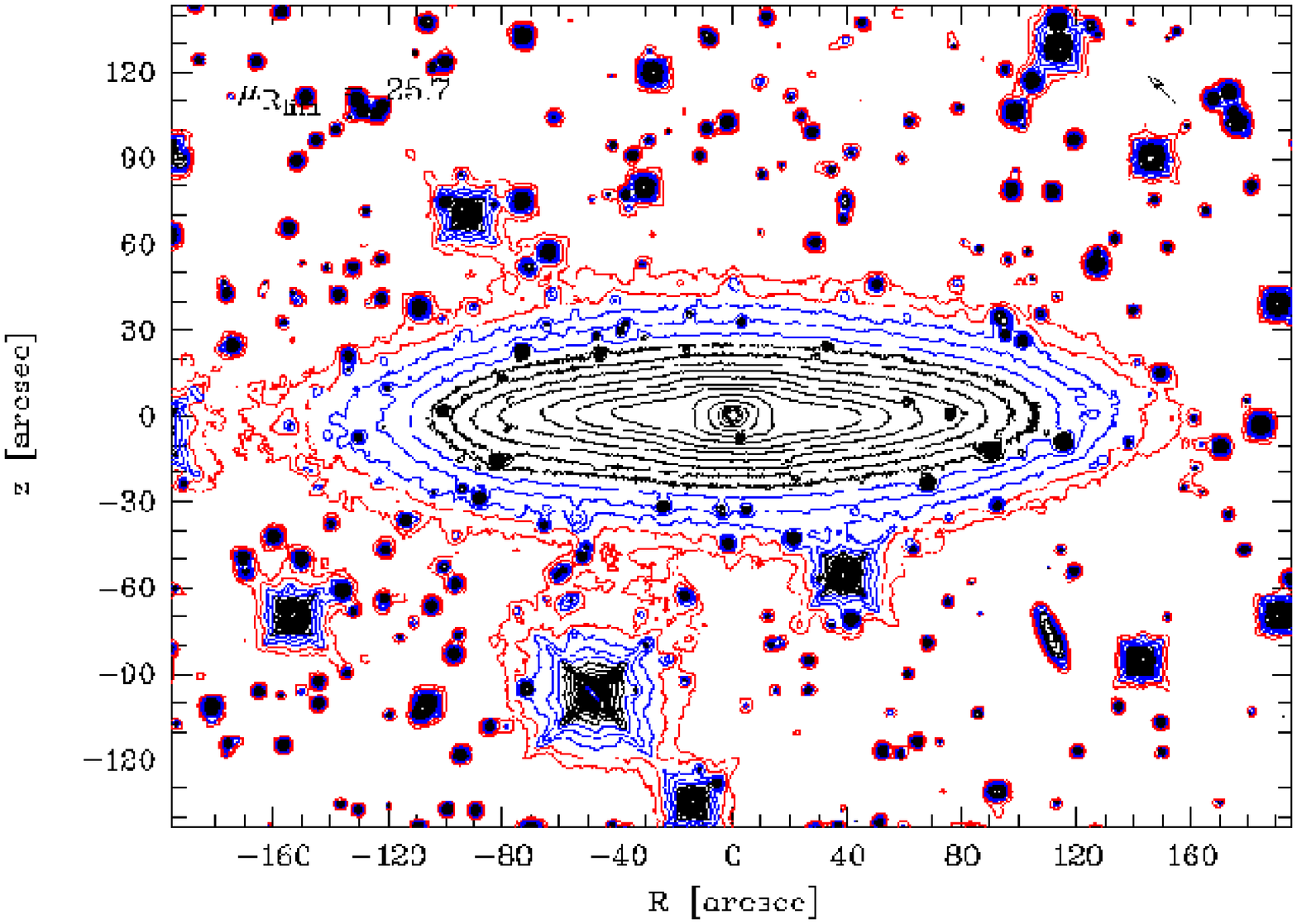}
\includegraphics[width=5.8cm,angle=270]{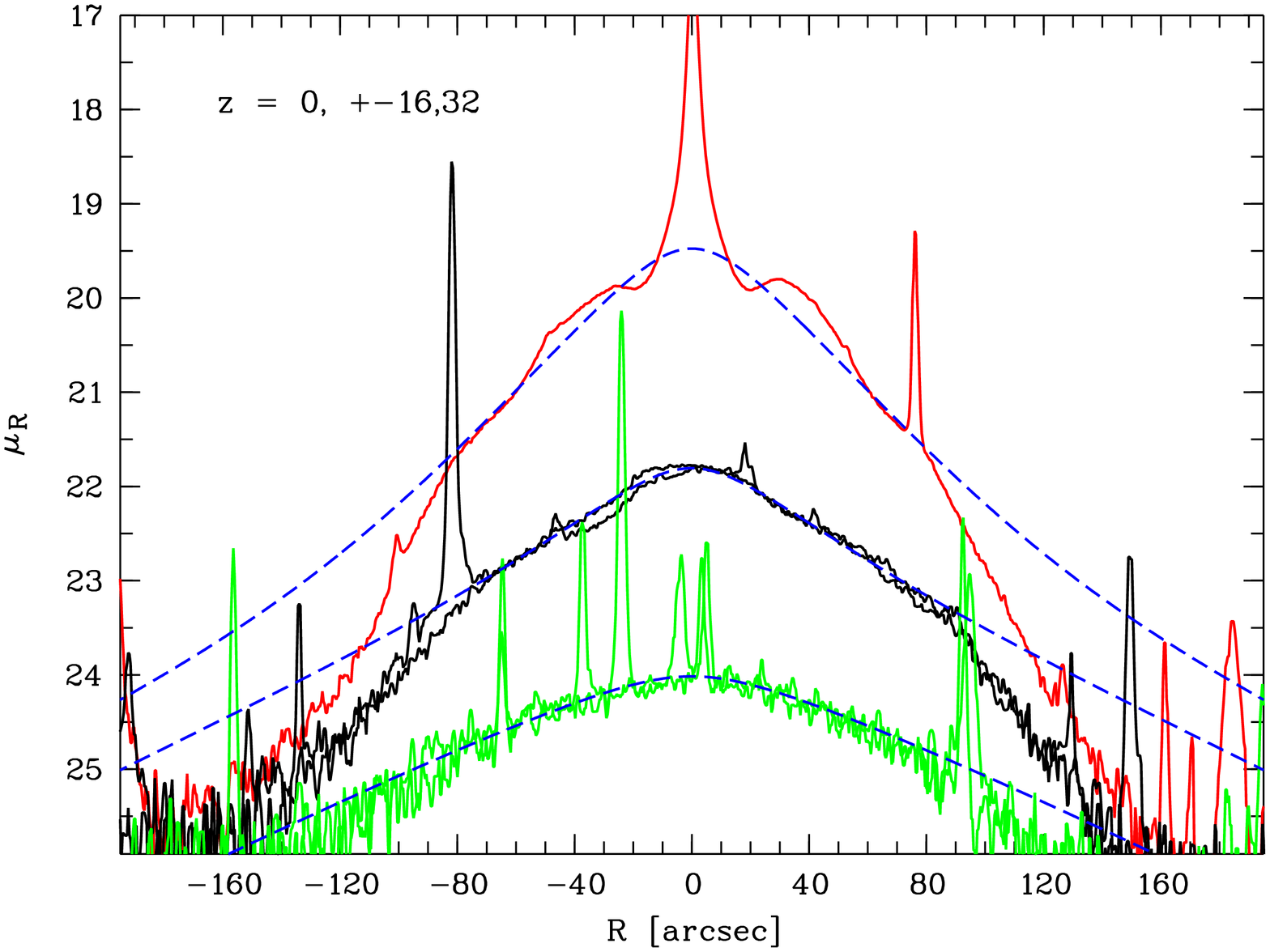}
\includegraphics[width=5.8cm,angle=270]{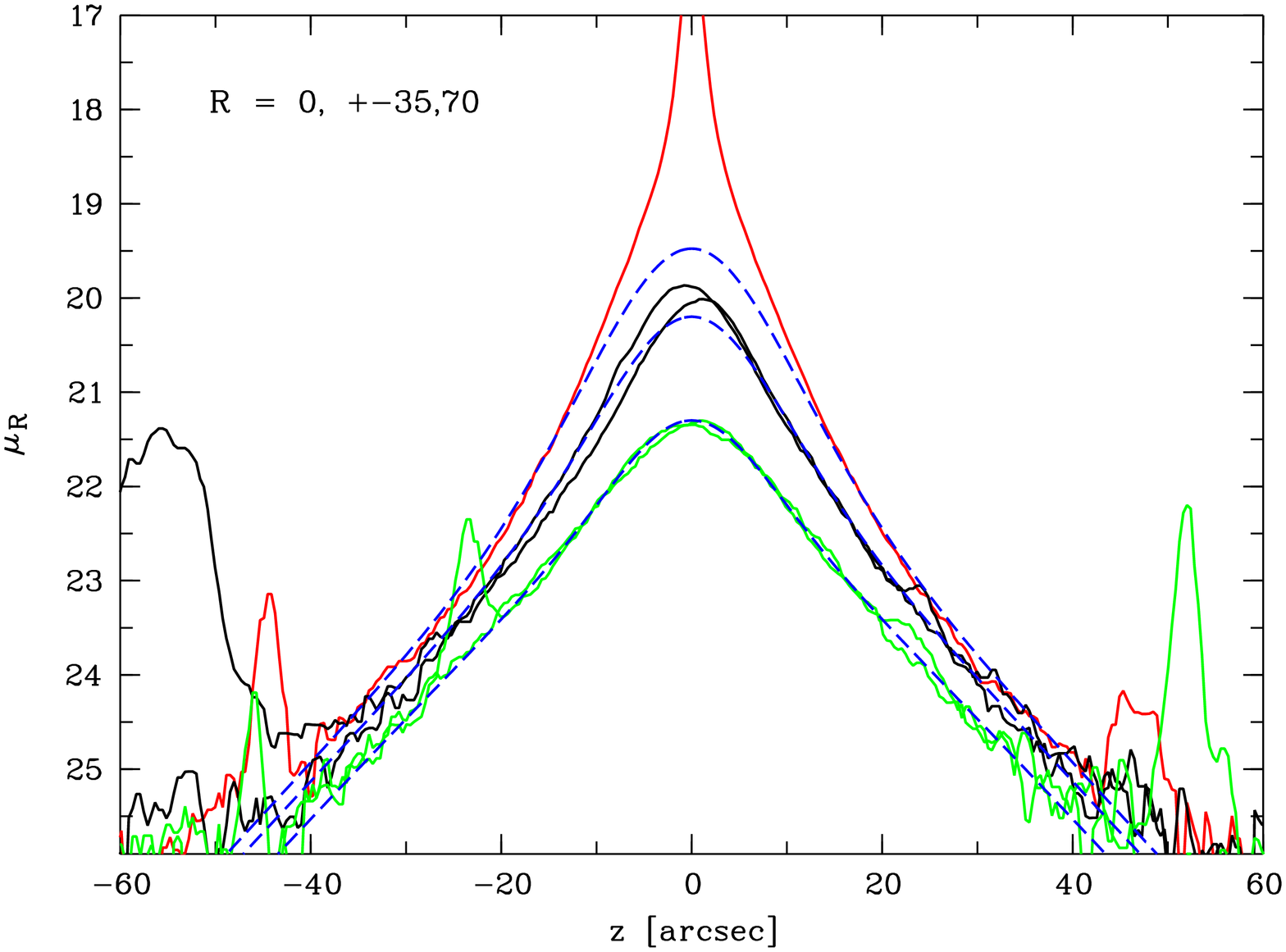}
\includegraphics[width=5.8cm,angle=270]{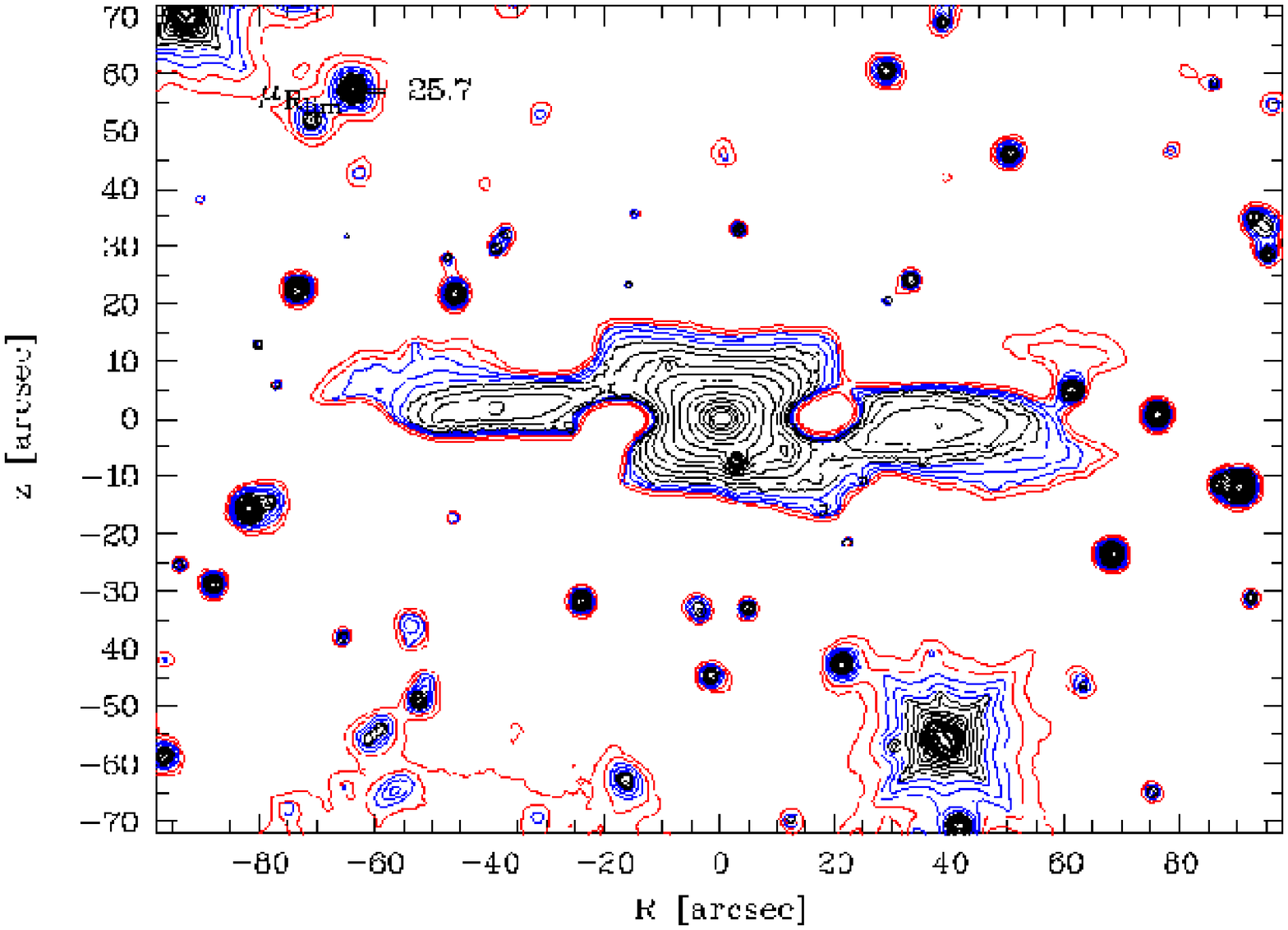}
\caption{NGC\,2310 R-band \label{n2310}}
\end{figure}

\begin{figure}
\includegraphics[width=5.8cm,angle=270]{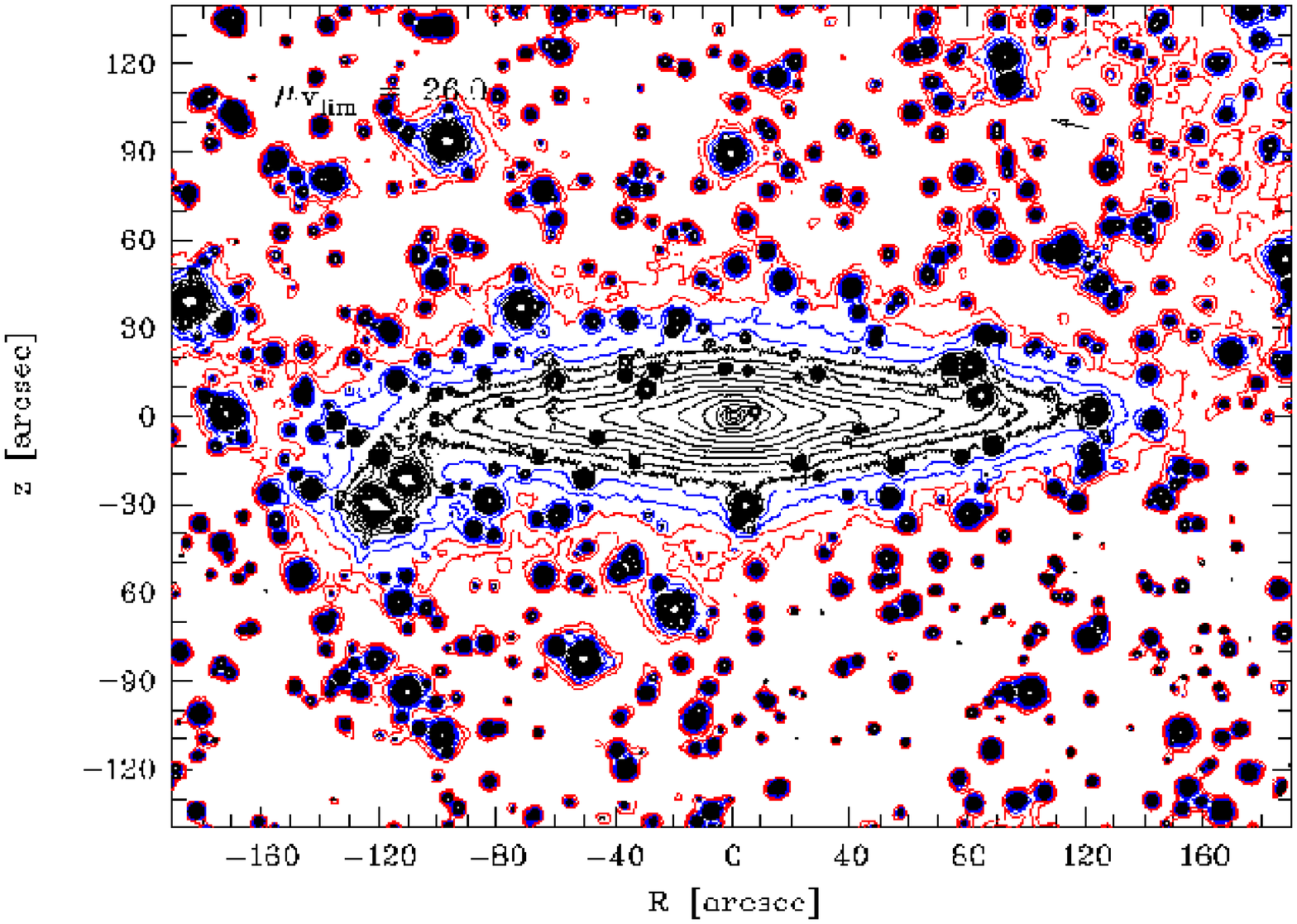}
\includegraphics[width=5.8cm,angle=270]{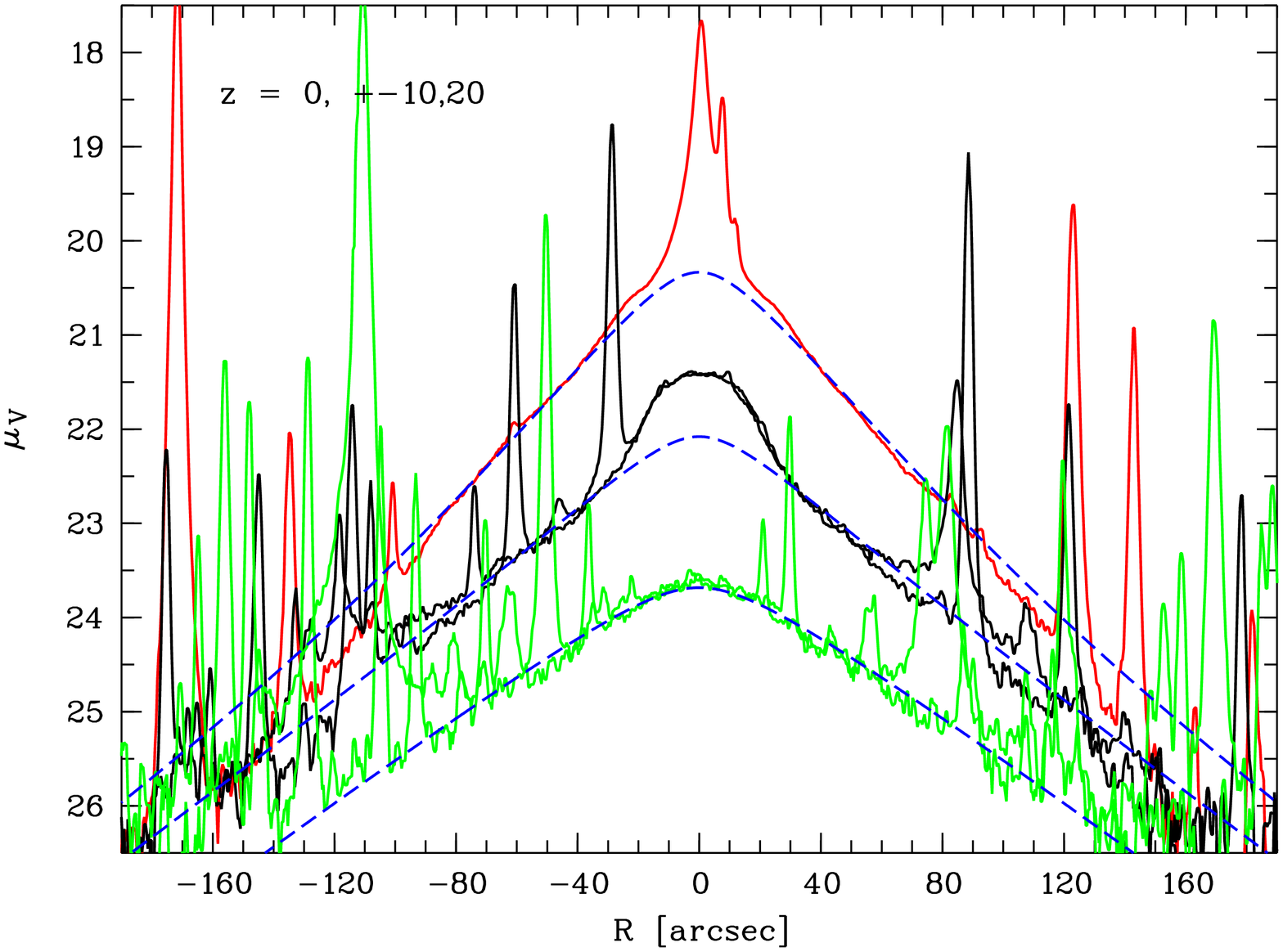}
\includegraphics[width=5.8cm,angle=270]{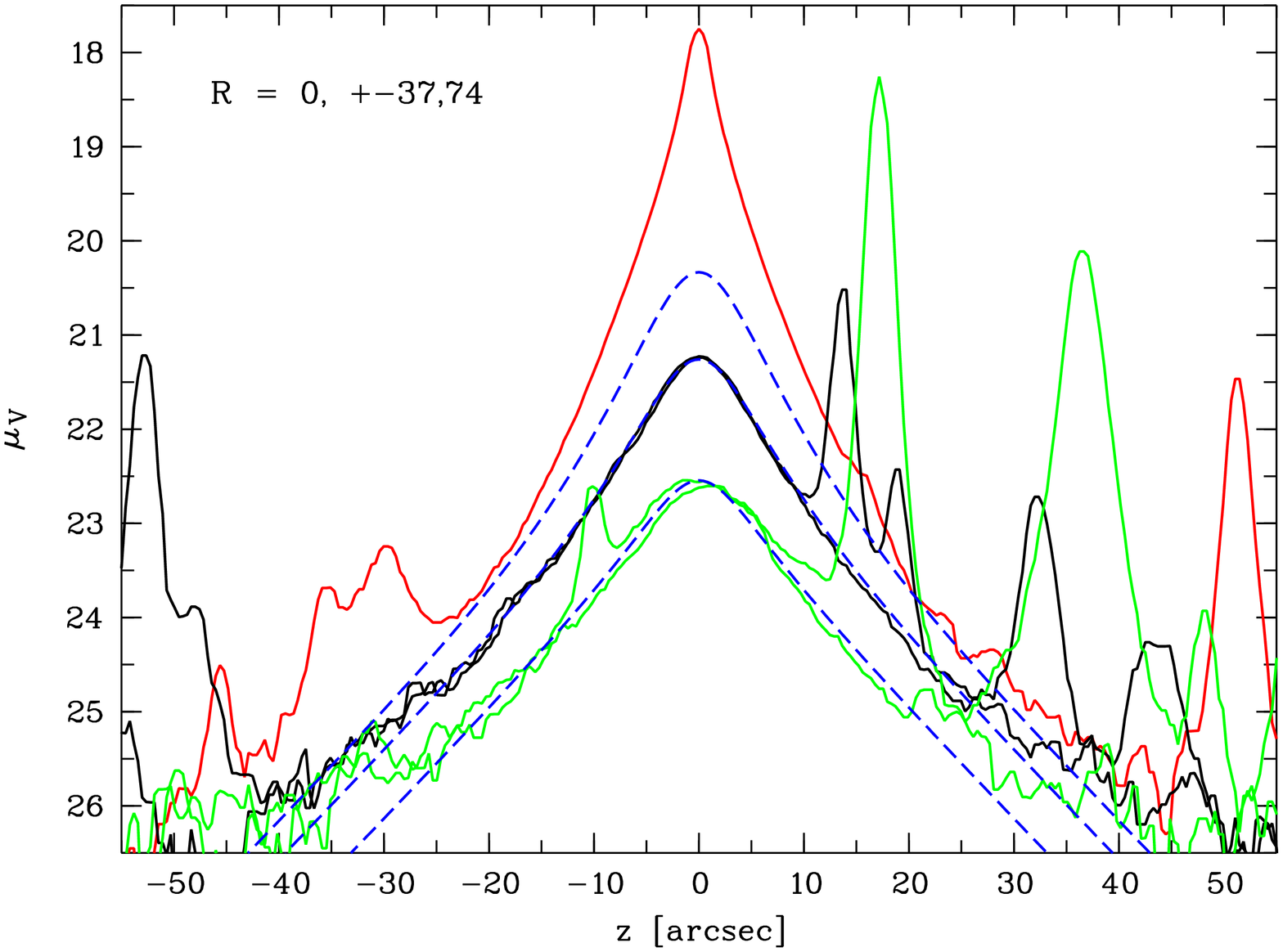}
\includegraphics[width=5.8cm,angle=270]{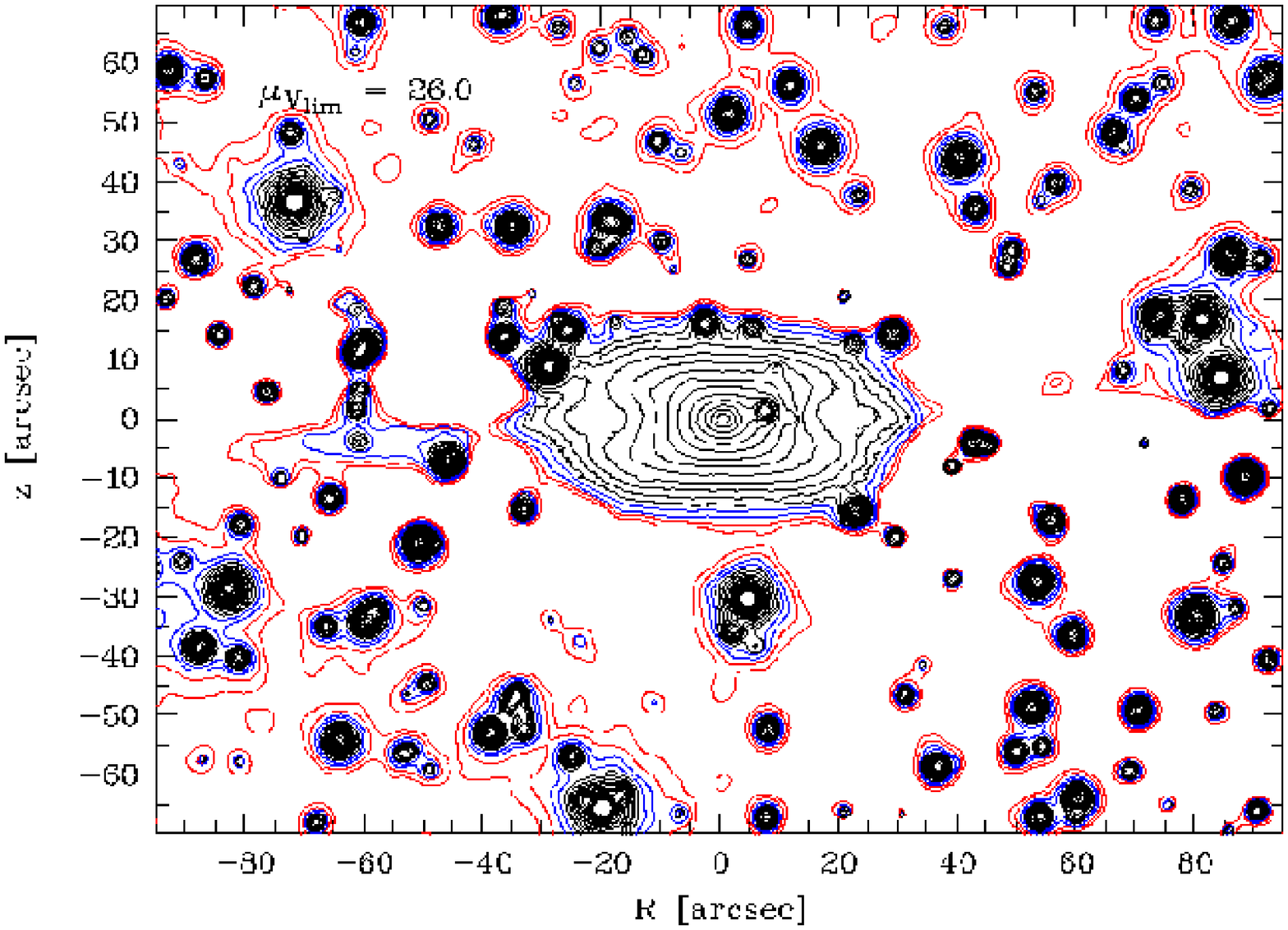}
\caption{ESO\,311-012 V-band \label{e311012} }
\end{figure}

\begin{figure}
\includegraphics[width=5.8cm,angle=270]{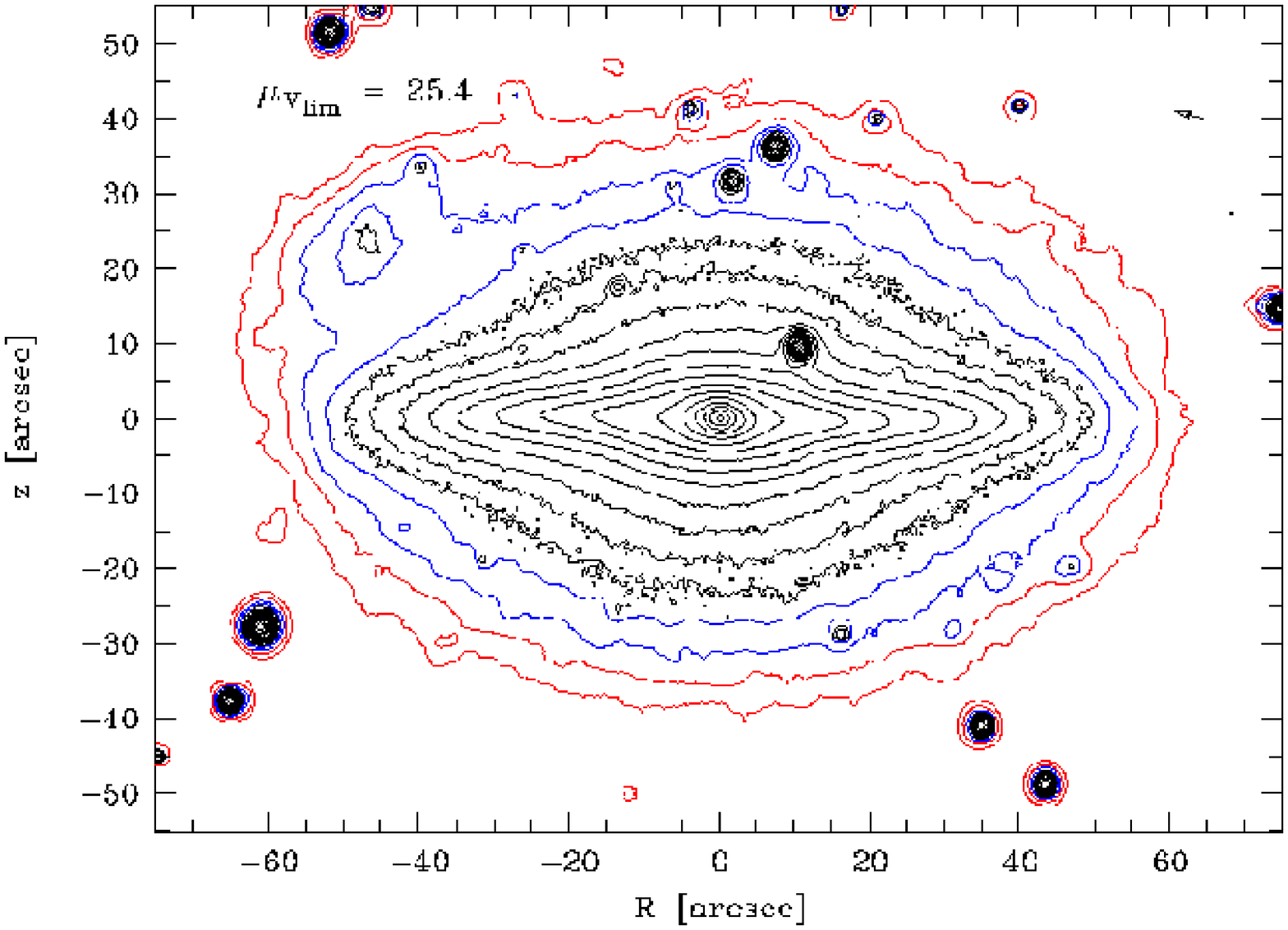}
\includegraphics[width=5.8cm,angle=270]{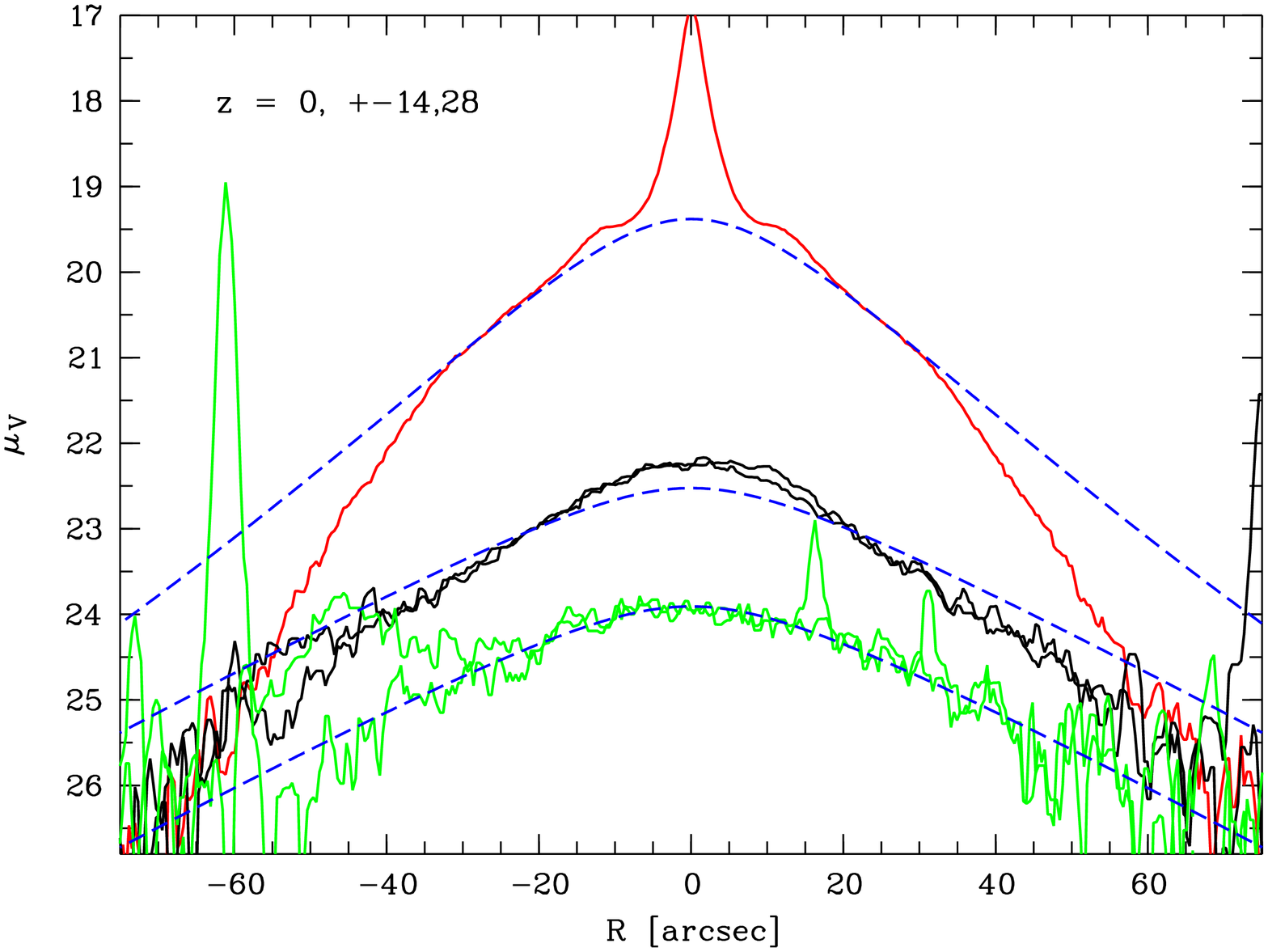}
\includegraphics[width=5.8cm,angle=270]{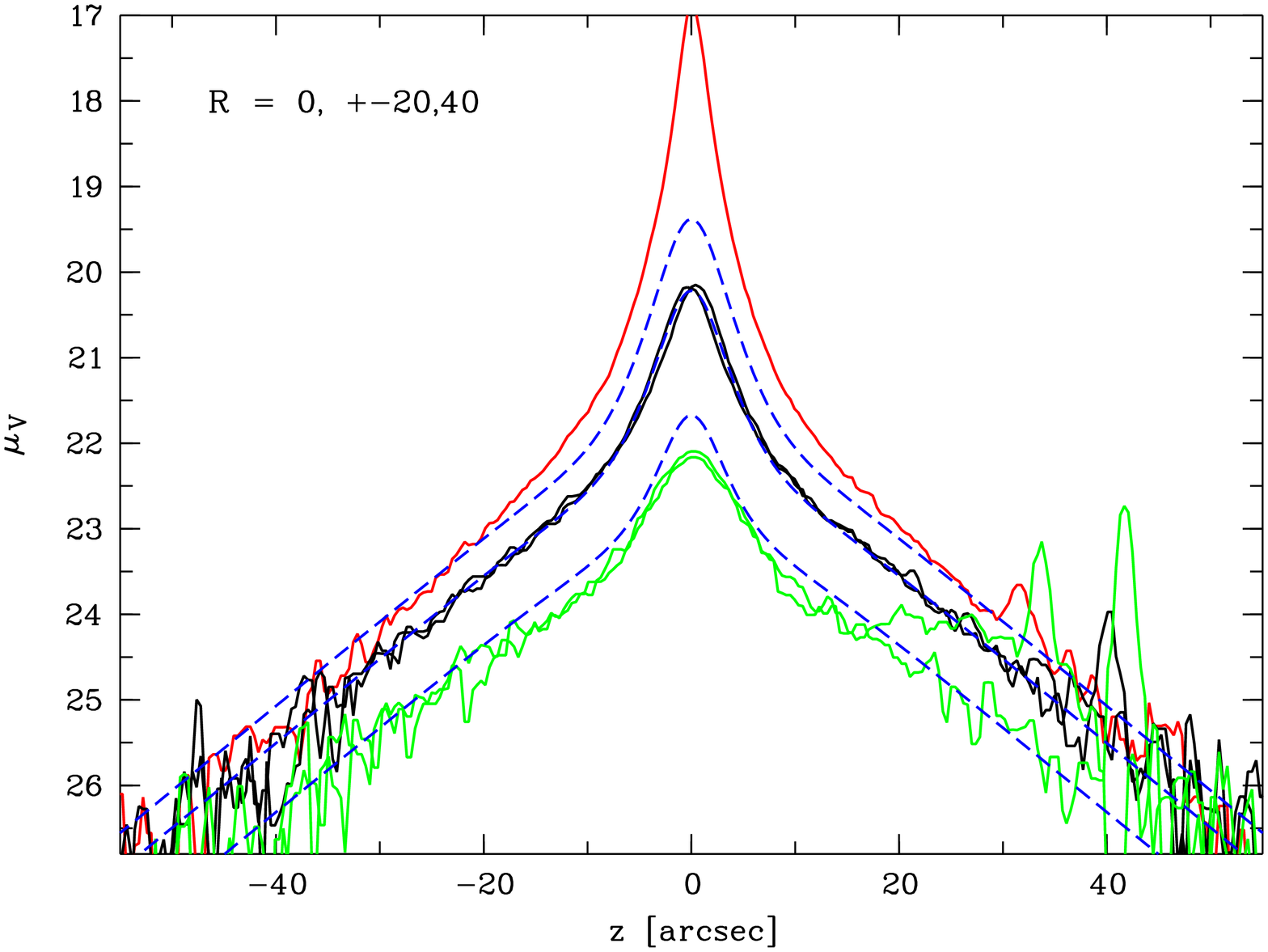}
\includegraphics[width=5.8cm,angle=270]{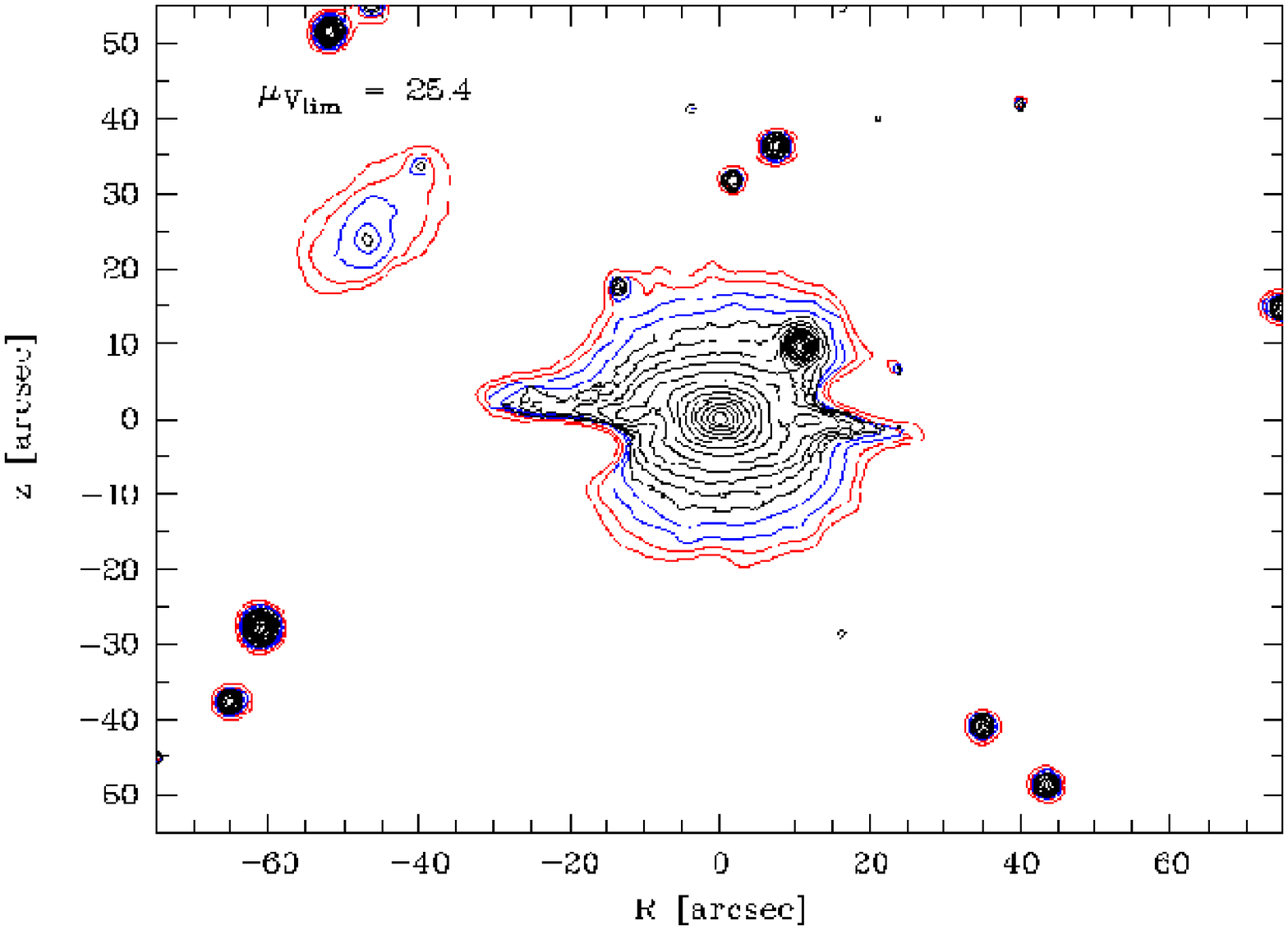}
\caption{NGC\,3564 V-band \label{n3564}}
\end{figure}

\begin{figure}
\includegraphics[width=5.8cm,angle=270]{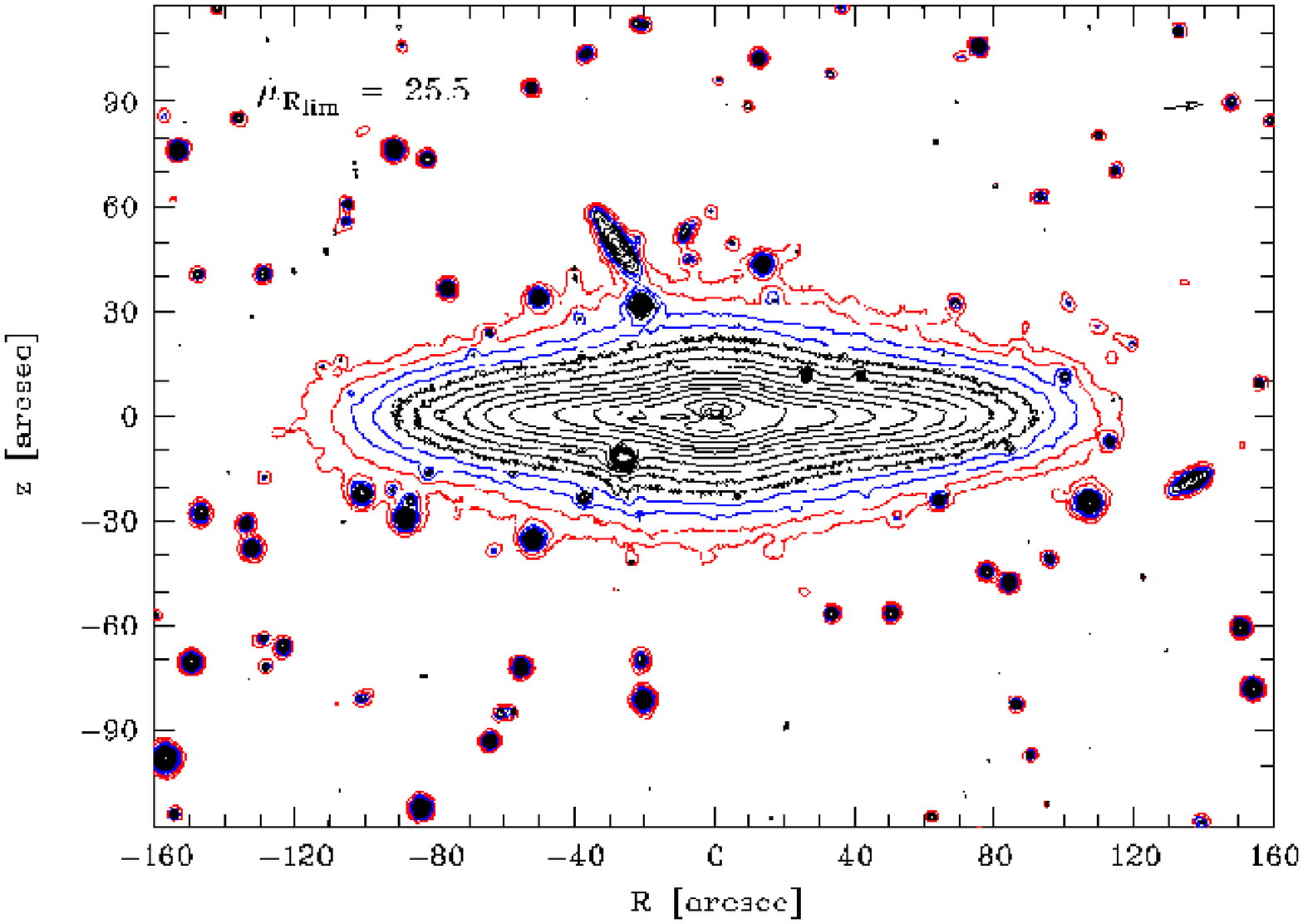}
\includegraphics[width=5.8cm,angle=270]{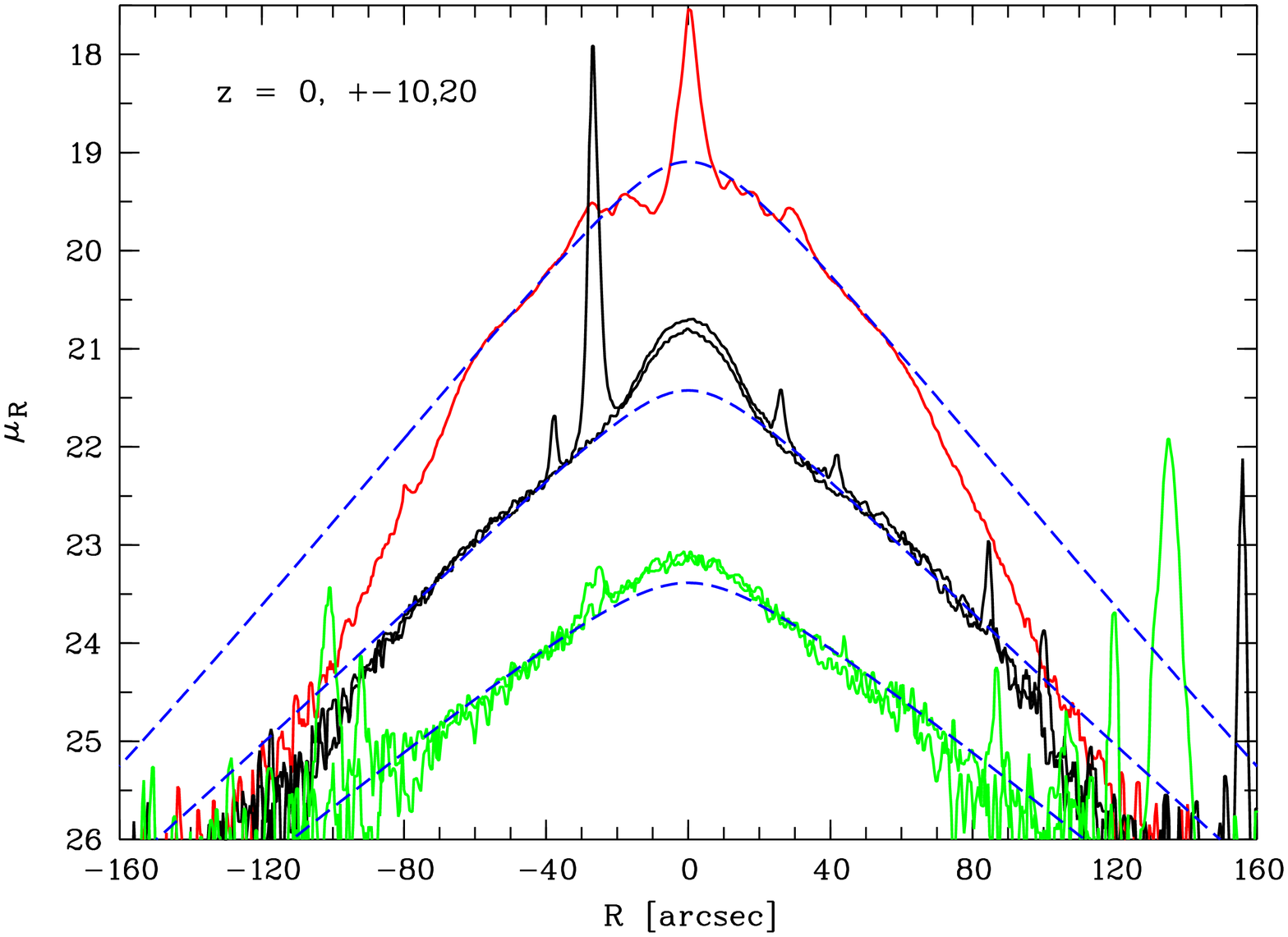}
\includegraphics[width=5.8cm,angle=270]{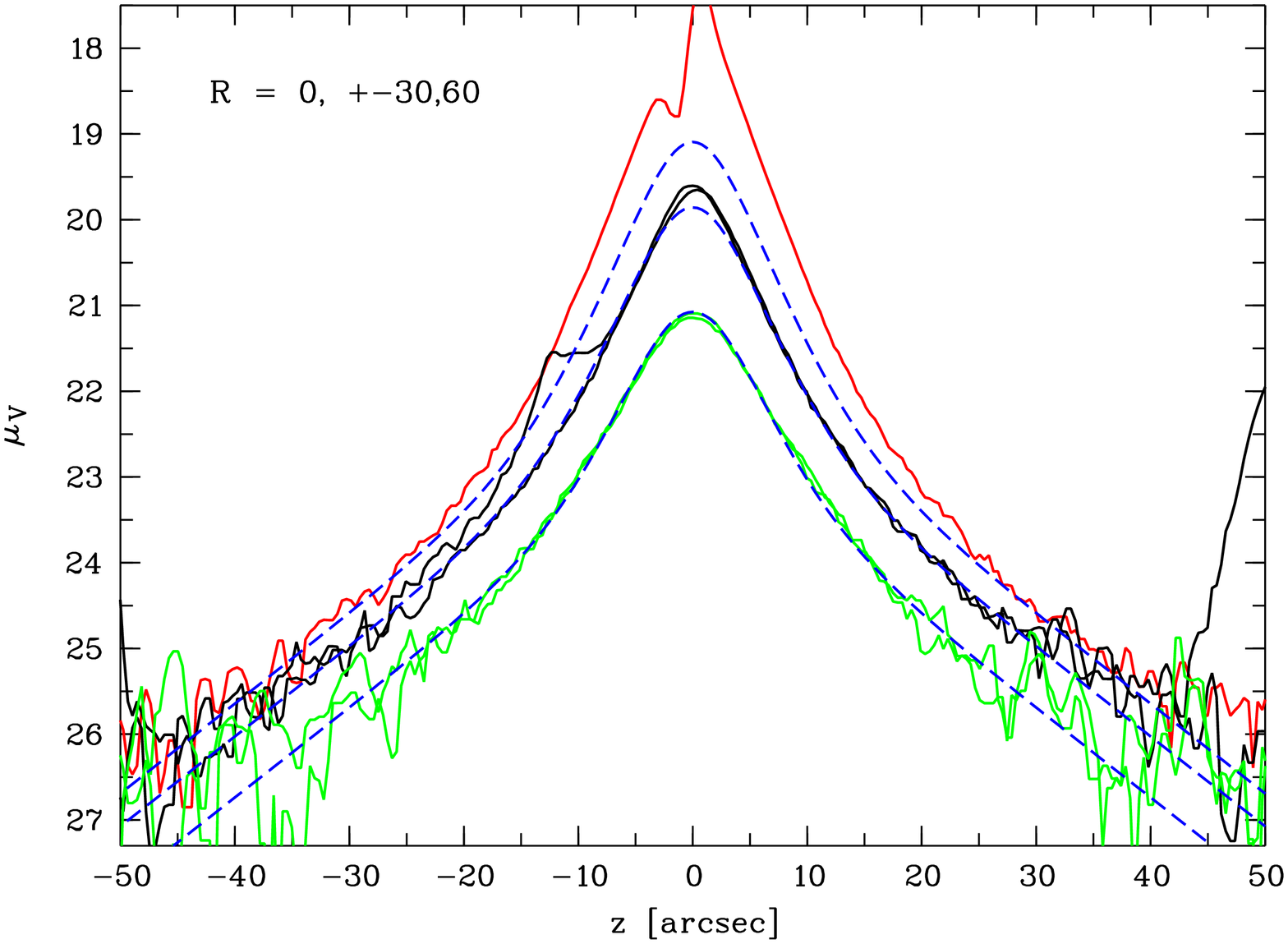}
\includegraphics[width=5.8cm,angle=270]{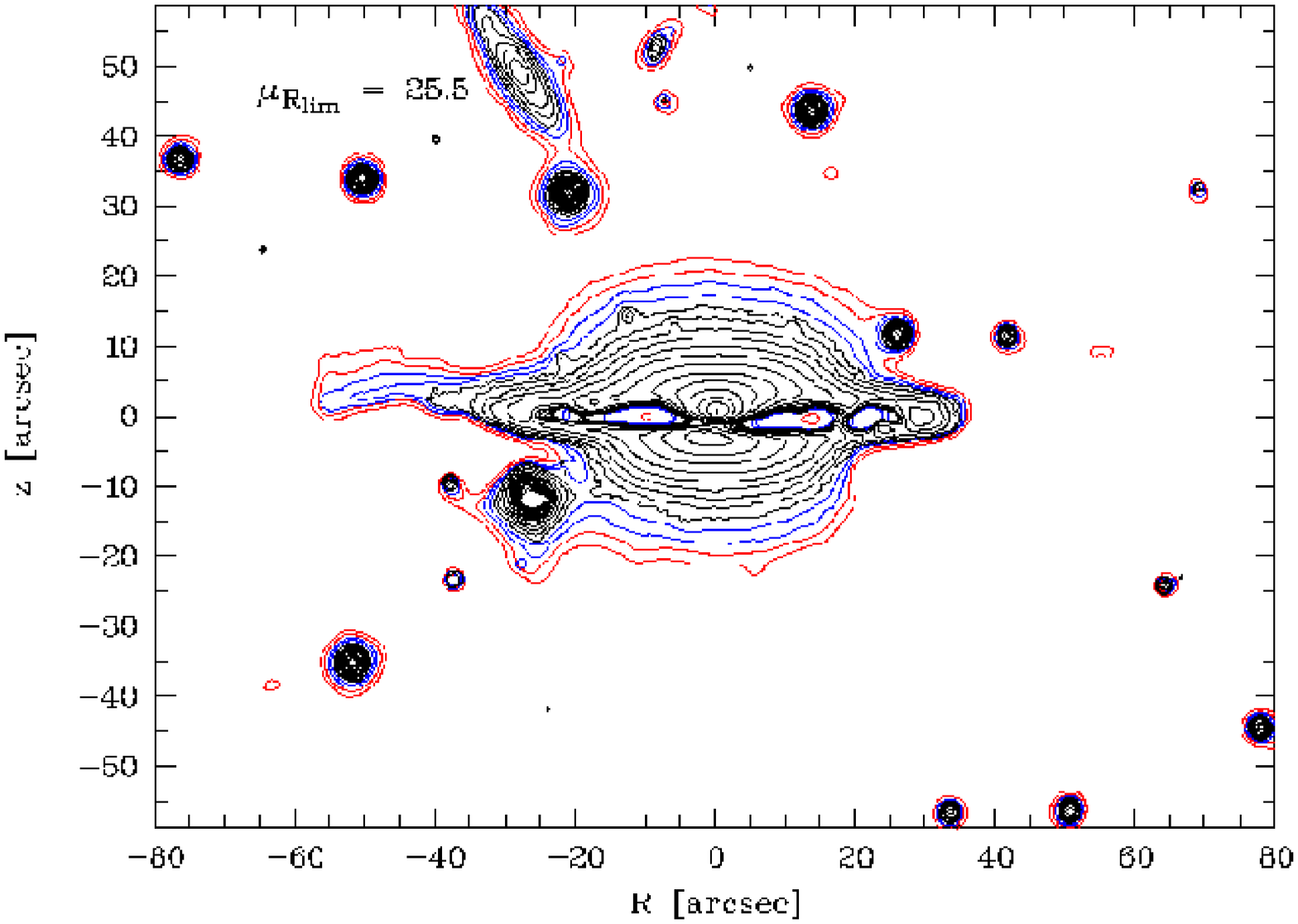}
\caption{NGC\,3957 V-band \label{n3957}}
\end{figure}

\begin{figure}
\includegraphics[width=5.8cm,angle=270]{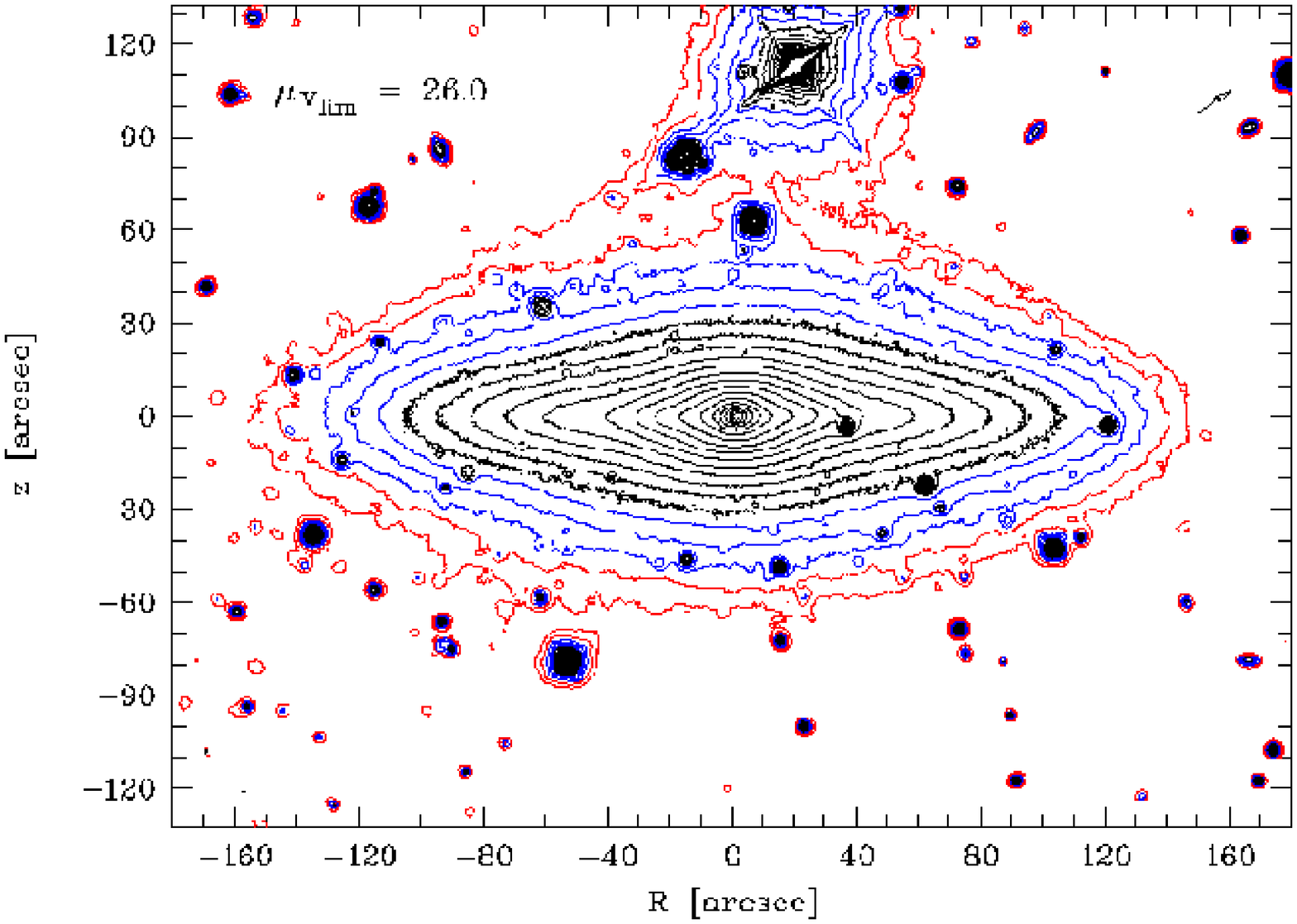}
\includegraphics[width=5.8cm,angle=270]{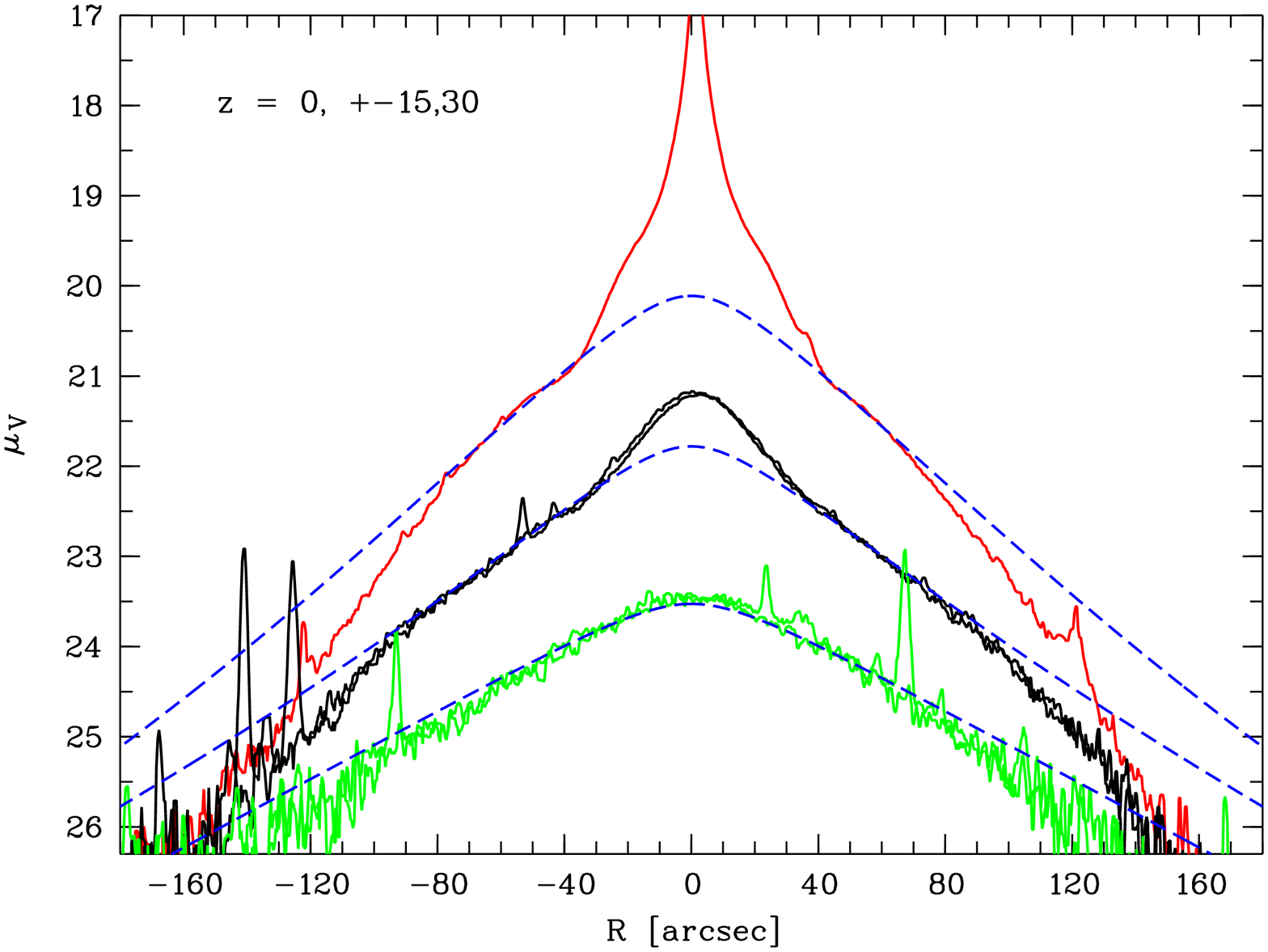}
\includegraphics[width=5.8cm,angle=270]{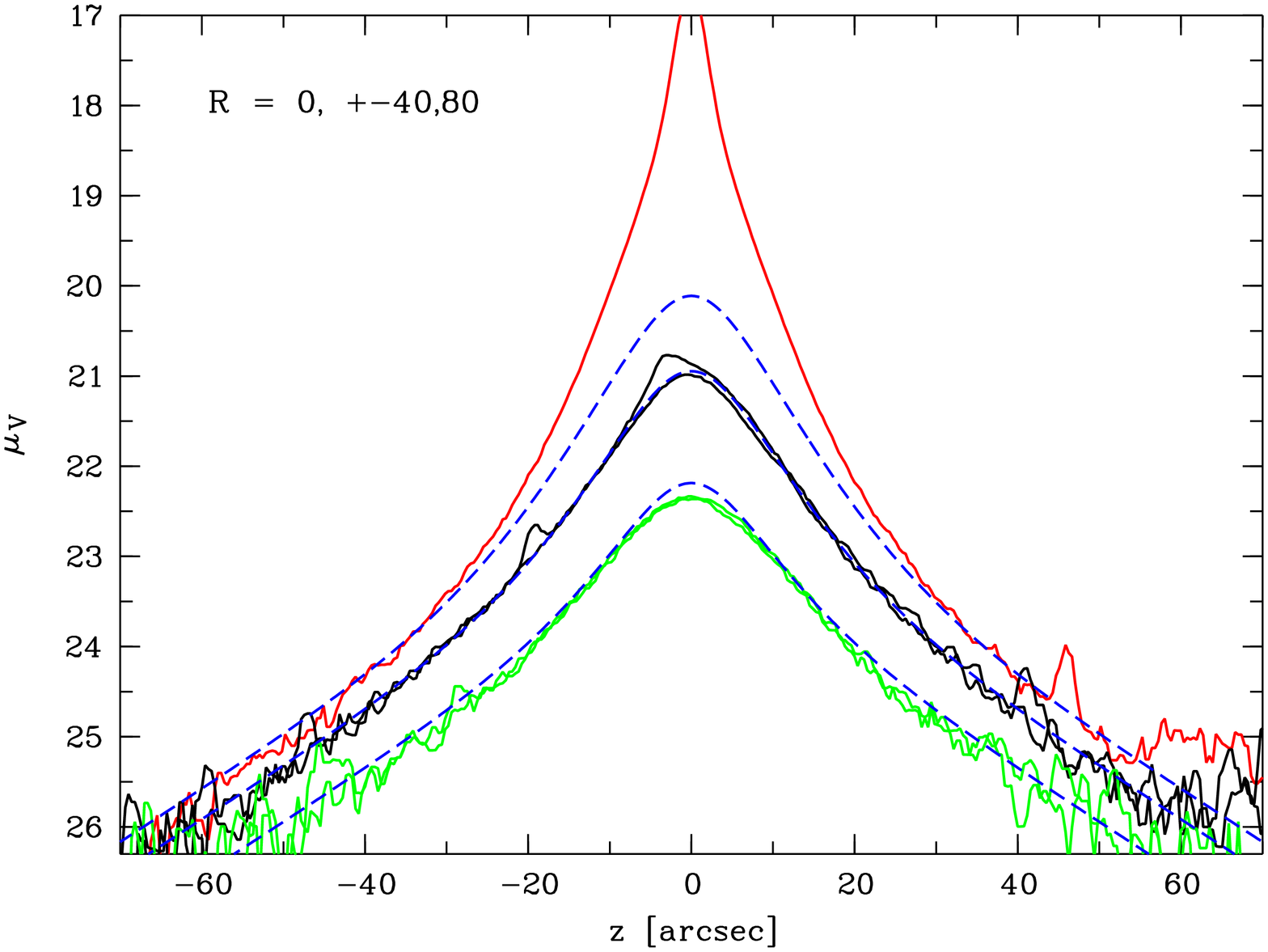}
\includegraphics[width=5.8cm,angle=270]{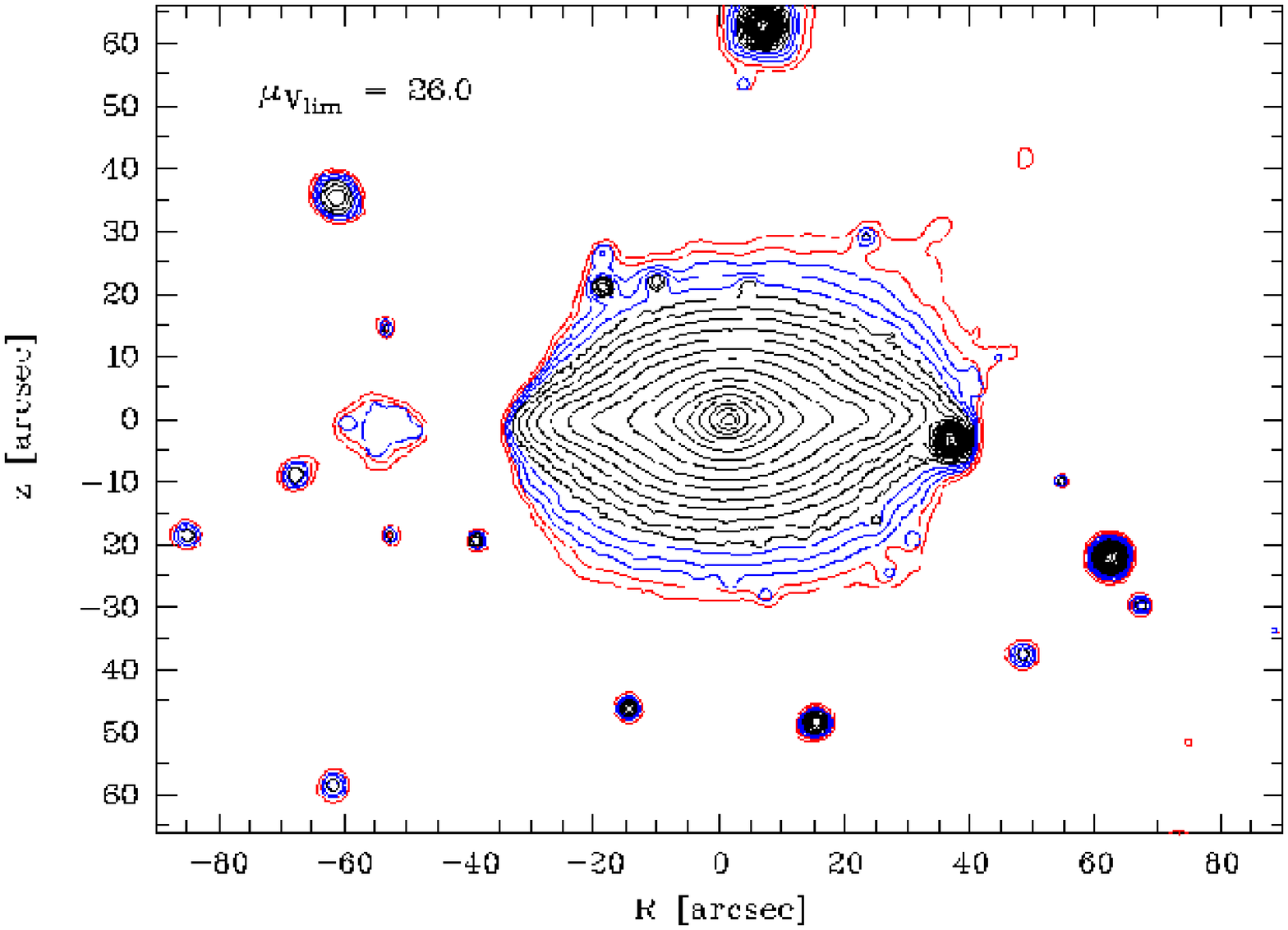}
\caption{NGC\,4179 V-band \label{n4179}}
\end{figure}

\begin{figure}
\includegraphics[width=5.8cm,angle=270]{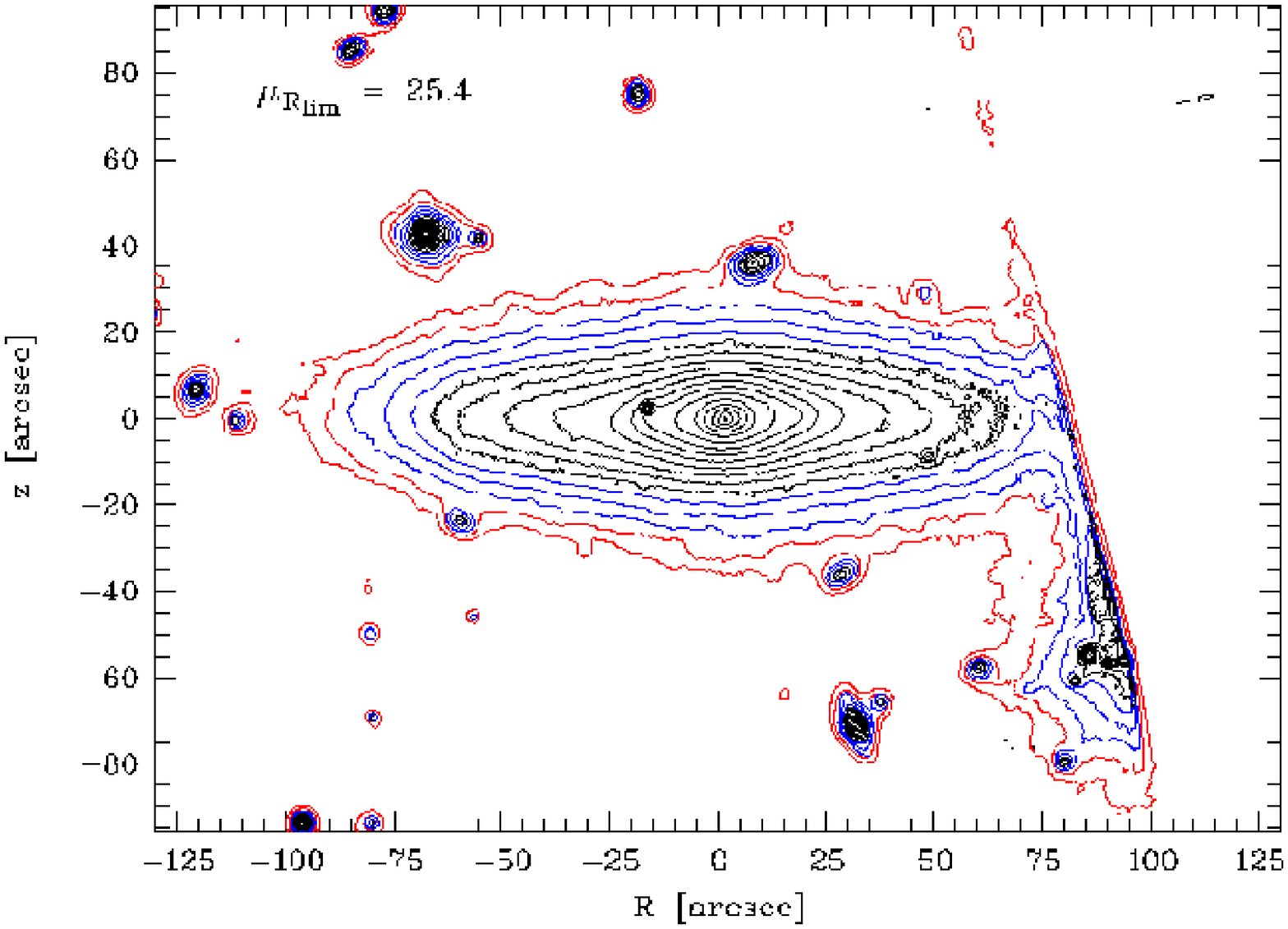}
\includegraphics[width=5.8cm,angle=270]{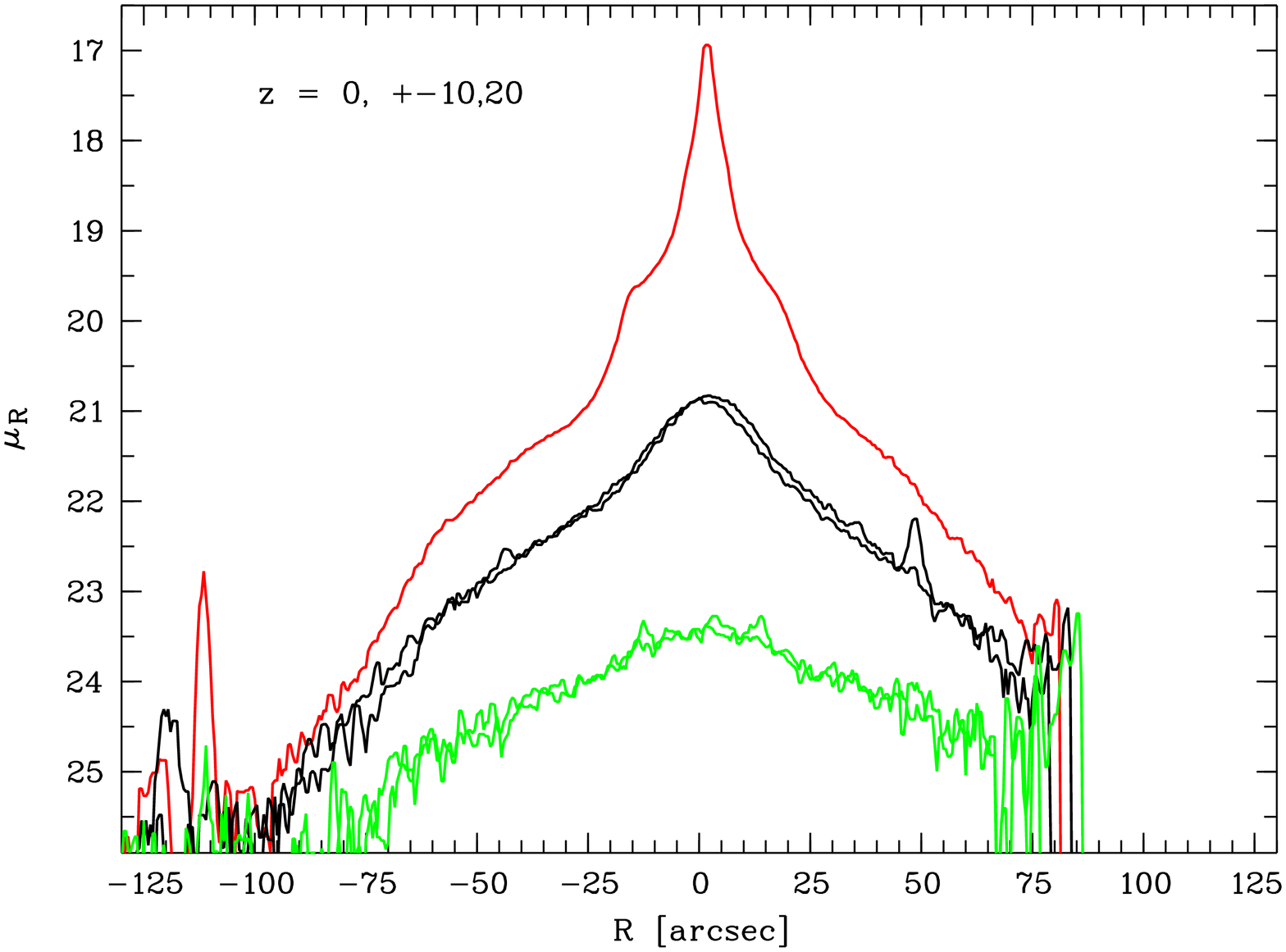}
\includegraphics[width=5.8cm,angle=270]{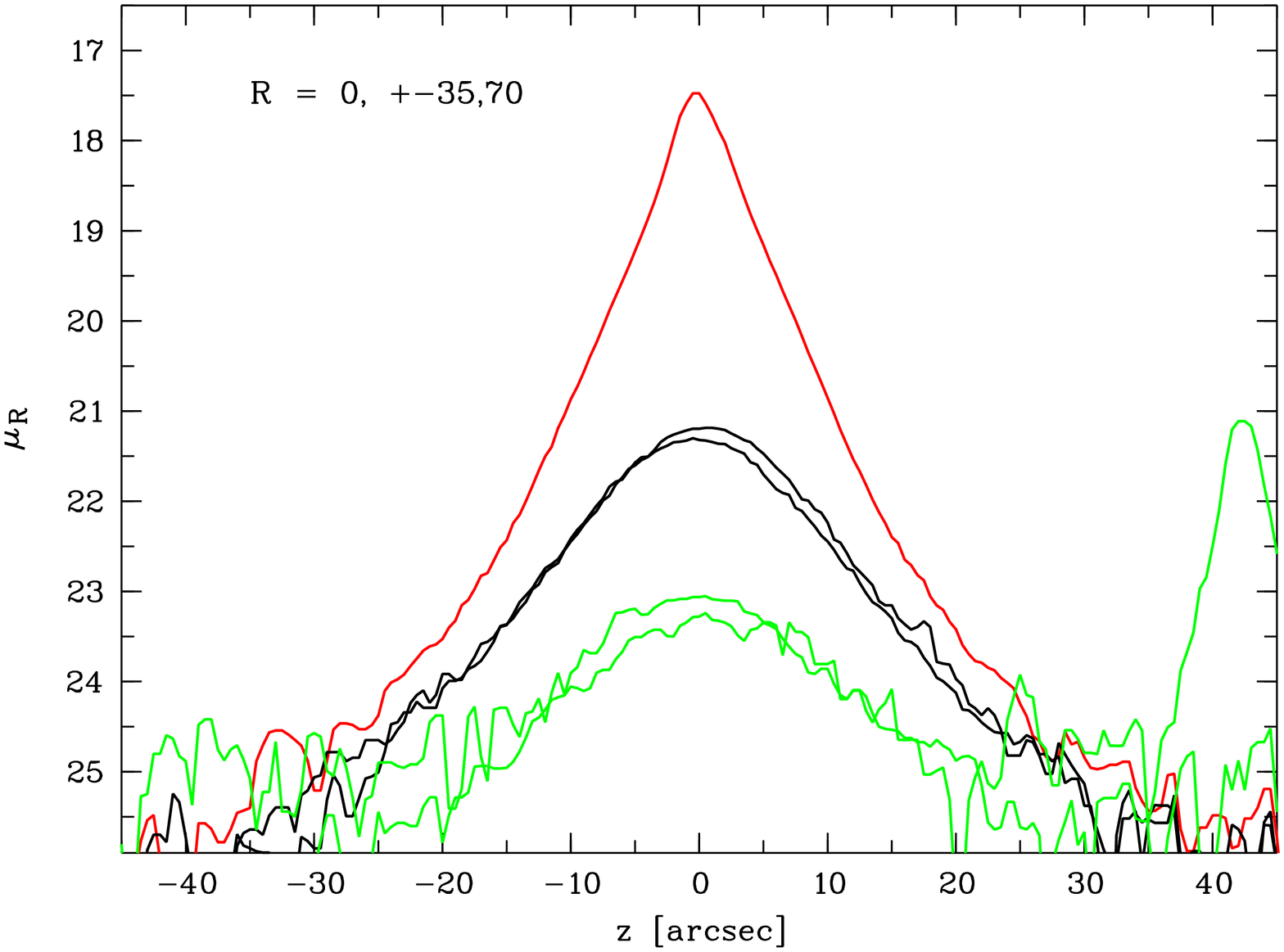}
\vspace*{5.85cm} 
\caption{NGC\,4521 R-band \label{n4521}}
\end{figure}

\begin{figure}
\includegraphics[width=5.8cm,angle=270]{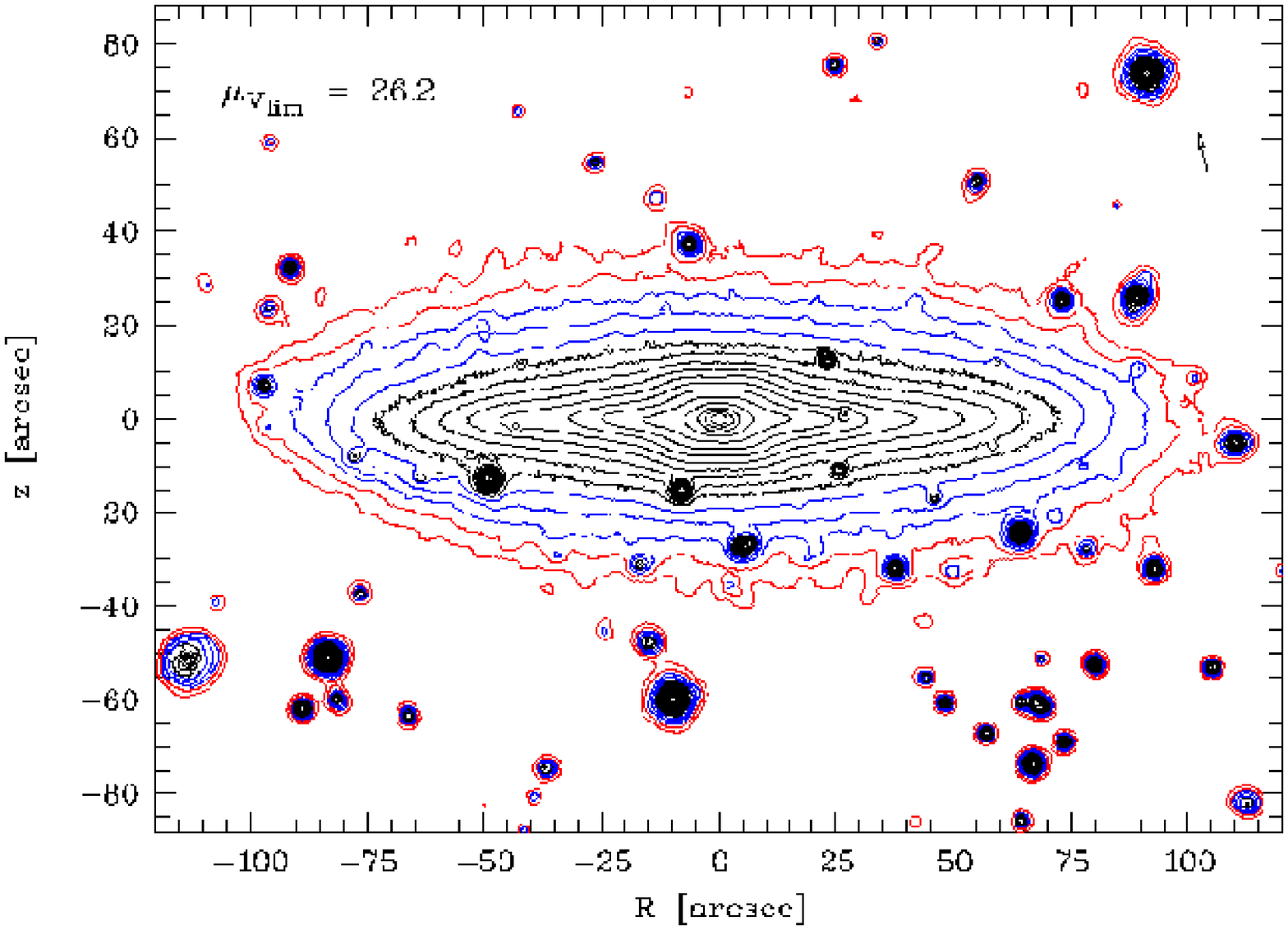}
\includegraphics[width=5.8cm,angle=270]{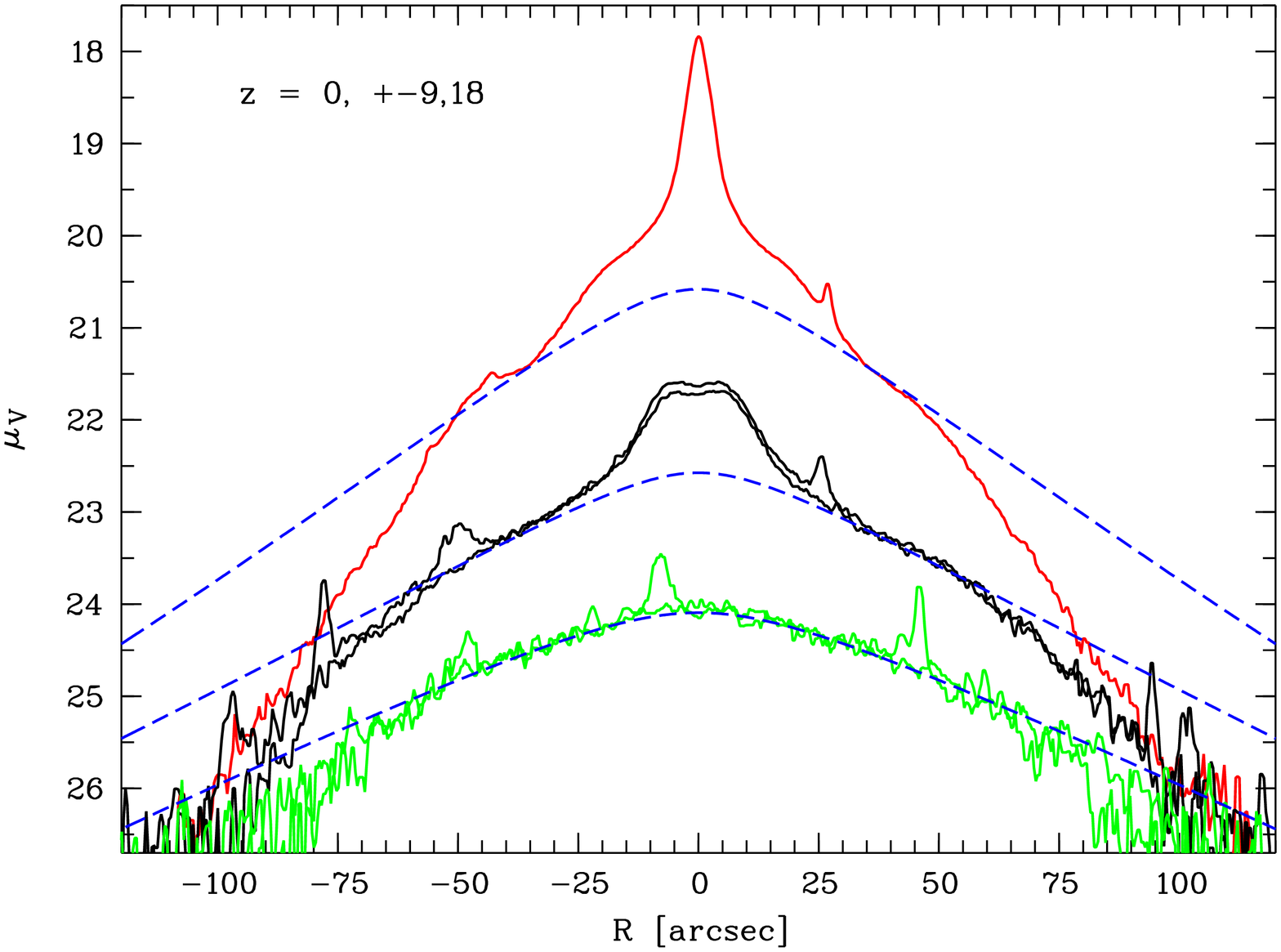}
\includegraphics[width=5.8cm,angle=270]{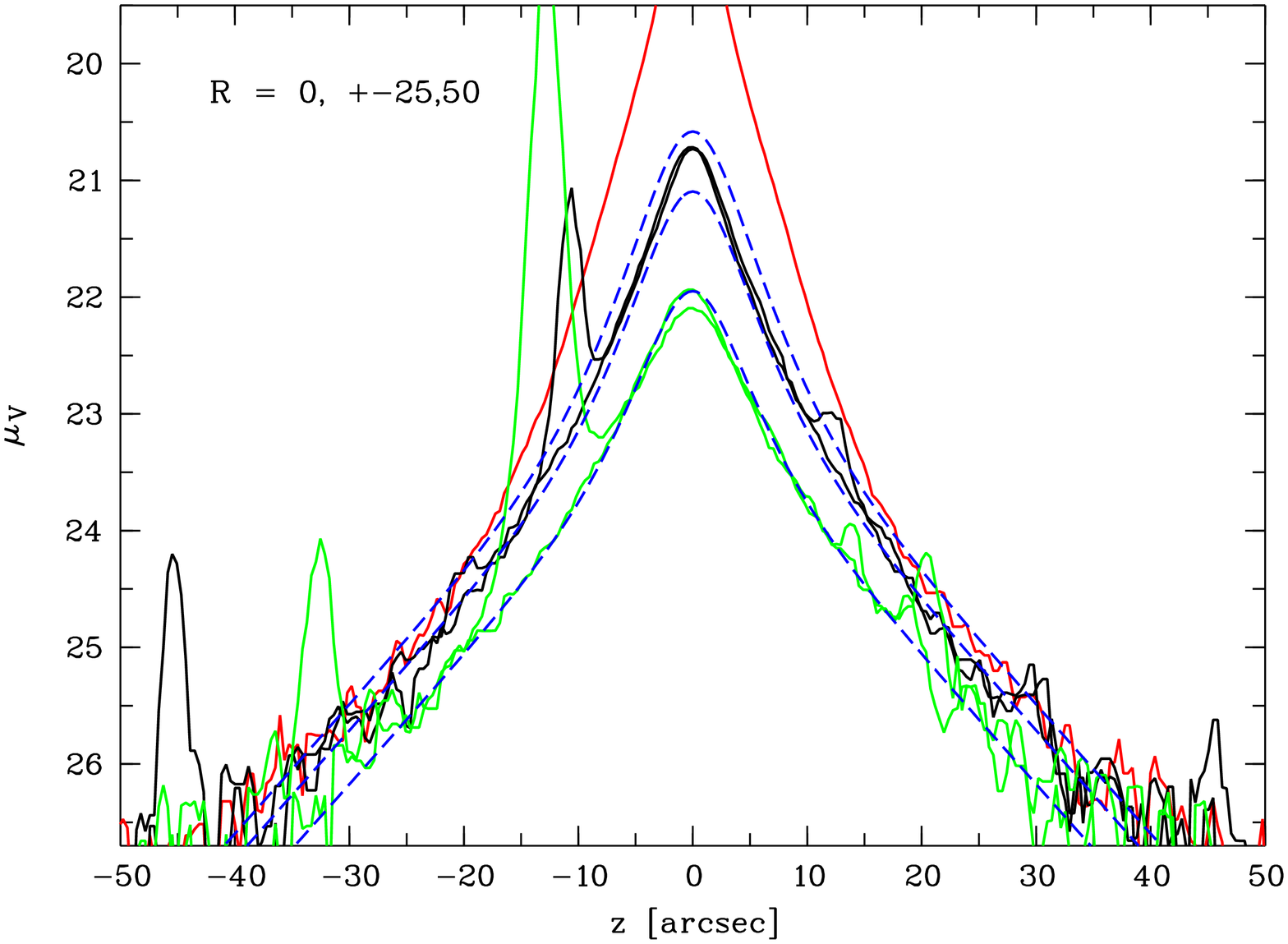}
\includegraphics[width=5.8cm,angle=270]{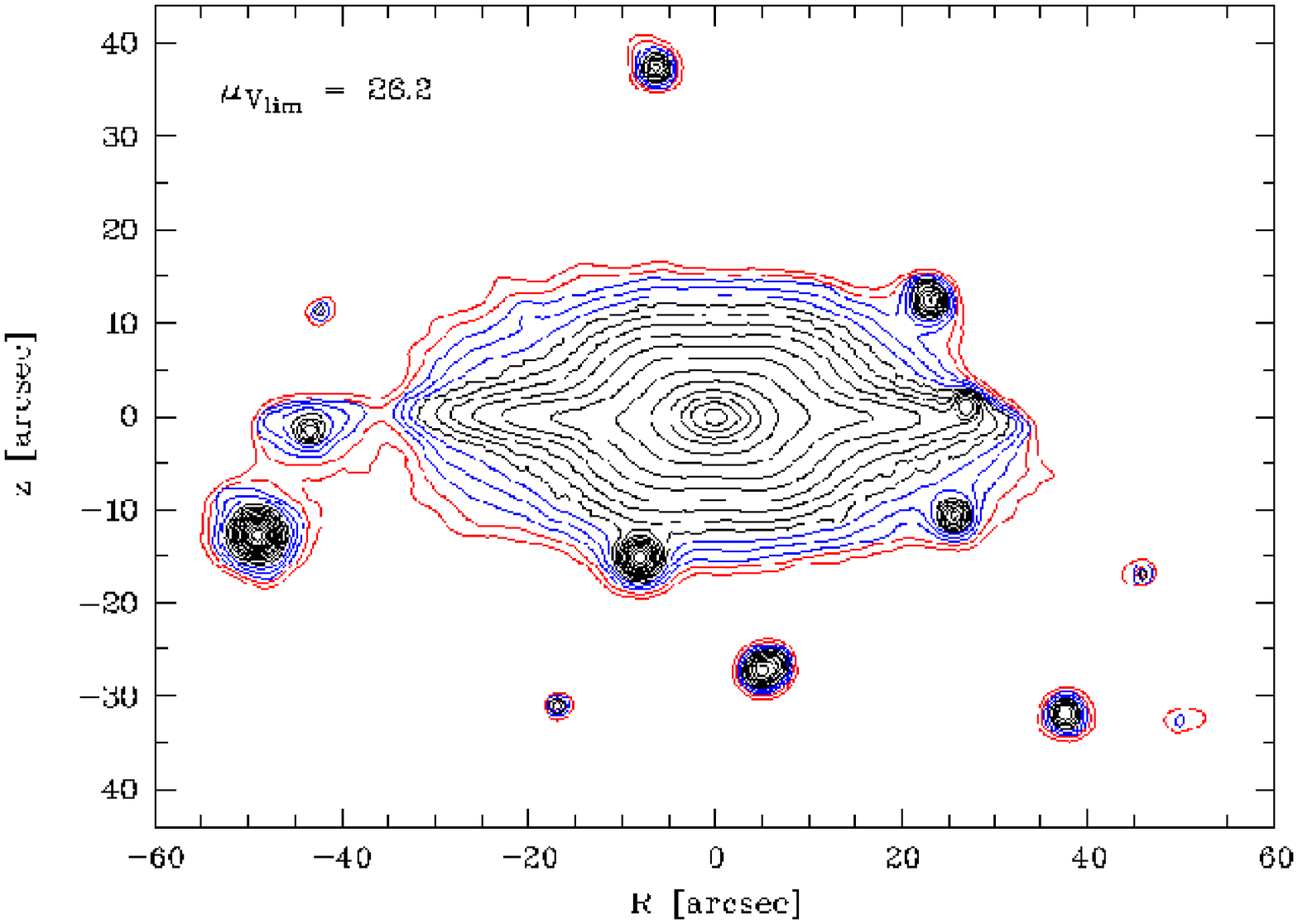}
\caption{NGC\,5047 V-band  \label{n5047}}
\end{figure}

{\bf NGC\,1596:} (\fig\ref{n1596}) 
A `normal' \s0 enclosed in a huge, spheroidal, outer envelope  
possibly triggered by interaction with the nearby physical companion 
(NGC\,1602). The difference from other \s0s is obvious from radial cuts. 
It is the only galaxy in which the profiles bend upwards 
(not caused by a background gradient) at larger 
galactocentric distances ($R\eq\pm115$\arcsec). This looks similar 
to an R$^{1/4}$ bulge component beginning to dominate the profile 
when extrapolating the curvature of the inner bulge towards the outer 
extension. However, the overall flatness of such an R$^{1/4}$ bulge 
would be unusual. In addition, the contours start to show substructure in 
the very outer parts (in contrast to the smooth contours of ellipticals). 
The inner light profiles are dominated by bulge and a possible 
inner disk component. The major axis shows several distinct parts: an inner 
bulge, a steep straight decline out to $R\!\approx\!\pm80$\arcsec, 
an asymmetric slightly shallower part out to  $R\eq-110$\arcsec\, 
and $R\eq+120$\arcsec, followed by an upwards bending profile 
an upbending profile. The SW-side is slightly more extended than 
the NE-side.
After the outer thick disk has been subtracted, 
the remaining profiles clearly 
resemble still the thin/thick disk behaviour (changing of slopes 
towards outer/higher radial cuts), therefore no consistent model 
could be derived. 
\noindent
{\bf NGC\,2310:} (\fig\ref{n2310}) exhibits a box-shaped bulge \cite[type 2 
according to][]{luett2000a}. The bulge/bar is rotated 
\mbox{$\approx\!0.5$\deg} against the outer disk, therefore we find a 
slightly tilted residual. The radial profiles show a complex structure 
with inner, small, very bright nuclear bulge component, and further 
out a significant dip in the profile (caused by a possible dust ring 
or mimicked by a subsequent stellar ring), sitting on top of a bar/ring-like 
structure $R\!\approx\!\pm50$\arcsec. There is a narrow disk-like structure 
$R\!\approx\!\pm82$\arcsec, here assumed to be 
the main underlying disk, finally showing a straight decline into 
the noise (NE-side disturbed by nearby star-halo). The best fitting model 
overestimates the inner disk component due to inner ring-like feature.
This feature is responsible for the apparent holes in the residual image, 
which shows the inner box-shaped bulge in addition to a large ring or bar.  
\noindent
{\bf ESO\,311-012:} (\fig\ref{e311012}) Box-shaped bulge \cite[type 2 
according to][]{luett2000a}. The galaxy position is close to the Galactic 
plane with many foreground stars overimposed. The bulge component is 
still visible on the minor axis high above the plane. The radial profile 
shows a bar-like structure at $R\eq\pm20$\arcsec, followed by an exponential 
disk component out to a break at $R\!\approx\!\pm83$\arcsec. 
The residual is a box-shaped bulge with the indication of a weak 
bar extending on both sides. 
\noindent
{\bf NGC\,3564:} (\fig\ref{n3564}) The galaxy is probably part of small 
group/cluster with at least two large, nearby physical companions
(NGC\,3568 and NGC\,3557). Taking into account the low surface brightness 
part it is different from the other \s0s. Out to 
$\mu_{\rm V}\eq23.4$\,\magsqarcsec it looks similar although with a 
more dominating bulge. The outer contours, on the other 
hand, reveal that NGC\,3564 is sitting in a huge, roughly
spherical stellar envelope, which has probably been built up by a merger 
with a smaller satellite galaxy still visible in this structure on the NE-side 
($(R,z)\eq-47\arcsec,+24$\arcsec, distance unknown).
The radial profile shows an inner bulge-dominated part, followed by a bar-like 
plateau $R\!\approx\!\pm10$\arcsec, sitting on top of an inner disk 
structure out to $R\!\approx\!\pm30$\arcsec which is followed by a 
straight decline into noise. 
The residual resembles a rather large, spheroidal inner bulge together
with a bar or ring-like structure extending to both sides. It nicely 
shows the possibly merging satellite. The thick disk component takes 
most of the light from the envelope and therefore the remaining bulge 
light is comparable to to that of the other galaxies. 
\noindent
{\bf NGC\,3957:} (\fig\ref{n3957}) Galaxy cluster in background. 
Nearly box-shaped bulge \cite[type 3 according to][]{luett2000a}. 
It is an \s0 galaxy with unusual, inner ($R\!\ltsim\!\pm30$\arcsec) dust lane. 
The bulge component is still visible on the minor axis high above the plane. 
The radial profiles show inner bulge, followed by a ring- or bar-like 
component at $R\!\approx\!\pm30$\arcsec (visible in cuts above the dust lane), 
with break at $R\!\approx\!\pm60$\arcsec\, and 
followed by a straight decline into the noise. The profile probably starts to 
bend outwards from $R\!\approx\!\pm105$\arcsec\  
($\mu_{\rm V}\gtsim24.3$\,\magsqarcsec) to $R\!\approx\!\pm150$\arcsec.
The residual resembles a nearly box-shaped bulge with an indication 
of a bar extending to both sides. The apparent holes are caused by 
absorption in the inner dust lane.  
\noindent
{\bf NGC\,4179:} (\fig\ref{n4179}) The inner bar/bulge component is rotated 
\mbox{$\approx\!0.5$\deg} against the outer disk. The major axis shows 
a bright inner bulge, a disk- or bar-like feature between 
$R\!\approx\!\pm40$\arcsec ($z\!\ltsim\!\pm5$\arcsec), a more 
shoulder-like decline out to $R\!\approx\!\pm80$\arcsec, here taken to
be the main underlying disk, followed by a straight decline into the noise.
The residual resembles a spheroidal bulge with an indication of
an inner disk or bar extending slightly to both sides.  
\noindent
{\bf NGC\,4521:} (\fig\ref{n4521}) 
The inner component is $\approx\!1$\deg rotated against the outer disk. 
The radial profile shows a bar- or ring-like structure between 
$R\!\approx\!\pm 20$\arcsec out to $z\!\ltsim\!\pm5$\arcsec; 
on top of a two-slope disk, inner part out to a break radius 
of $R\!\approx\!-60$\arcsec, followed by a steeper straight decline 
out to $R\!\approx\!-77$\arcsec and a slightly upwards bending part at the 
very edge (perhaps flatfield problem).
The only galaxy with inconsistent thick/thin disk fitting without obvious 
reason (\cf NGC\,1596). All trials, such as changing vertical fitting 
function, have failed. When various thick disk models have been subtracted, 
the resulting thin disks always contributes too much light in the 
outer/higher z profiles dominated by the thick disk.  
Because of this the following thick disk is flattened, producing an 
unstable situation, perhaps caused due to lower image quality, proximity 
to chip edge, and disturbed by bright star on the NW-side. 
In principle, the galaxy exhibits the same features in radial/vertical 
cuts as our other \s0s. 
\noindent
{\bf NGC\,5047:} (\fig\ref{n5047}) Peanut-shaped bulge \cite[type 1 
according to][]{luett2000a}.
There are many companions around associated with the foreground NGC\,5044 
group of galaxies ($\Delta v\!\approx\!3600$\,\kms). Outer disk has a 
slightly s-shaped warp, on the SW-side towards N, on the NE-side towards S. 
The inner disk part is slightly off-centered to the outer component. The 
radial profiles show a bar-, or ring-like feature between 
$R\!\approx\!\pm 30$\arcsec out to $z\!\ltsim\!\pm3$\arcsec, on top of 
a shoulder-like inner disk out to $R\!\approx\!\pm 50$\arcsec, followed 
by a straight decline out to $\mu_{\rm V}\gtsim25.0$\,\magsqarcsec. 
The residual shows a nice peanut-shaped bulge with a thin bar-like
component. The ratio of the bar length to the maximum of the peanut
distortion closely matches simulations explaining the peanut by dynamical
processes in the bar potential \cite[cf.][]{luett2000b}.\\[-0.8cm]   
%
%
\section{Isophote maps, radial/vertical profiles, and residual images}
\label{app}
The following figures show in the {\sl upper panel} an isophote map 
of each galaxy rotated to the major-axis. 
A small arrow in the upper right corner indicates the North.
The magnitude of the outer contour ($\mu_{\rm lim}$, defined by 
a 3$\sigma$ criterion of the background)
is indicated in each plot. The consecutive contours are equally spaced 
by 0.5\,mag. 
The contour lines are plotted with increasing smoothing 
towards the outer parts. For the inner contours out to where the noise 
begins to increase no smoothing was applied. The following 2-3 contours 
are smoothed by replacing each pixel by the mean of $3^2$-pixels "around" 
the central pixel. For the outer two contours this smoothing is increased 
to $5^2$-pixels.
The {\sl second panel} displays the major-axis surface brightness profile 
(top) and two parallel radial profiles each above and below the major axis. 
The exact $z$-positions for the plotted profiles are indicated in the upper 
left corner of the plot.  
For a consistent thin/thick disk decomposition the model 
profiles are overplotted as dashed lines. 
The {\sl third panel} displays the minor-axis surface 
brightness profile (top) and two parallel vertical profiles 
on either side of the minor-axis. Again the exact $R$-positions 
for each cut are marked in the upper left corner and 
the best model is overplotted with  dashed lines.
The {\sl lower panel} is the isophote map of the residual image 
after subtracting the best fitting thin/thick disk model. The 
magnitude of the outer contour ($\mu_{\rm lim}$) is again indicated 
in each plot and the following contours are equally spaced by 0.5\,mag.   
%
%
%

\end{document}